\RequirePackage[hyphens]{url}
\documentclass[letterpaper,twocolumn,10pt]{article}
\usepackage{usenix2019_v3}

\usepackage{tikz}
\usepackage{amsmath}
\usepackage{enumitem}
\usepackage{caption}
\usepackage{afterpage}

\usepackage{tcolorbox}
\tcbuselibrary{breakable}

\usepackage{titling}
\usepackage{filecontents}
\usepackage[subtle,bibnotes,leading,mathdisplays]{savetrees}

\newlength{\bibitemsep}
\newlength{\bibparskip}
\let\oldthebibliography\thebibliography
\renewcommand\thebibliography[1]{%
	\oldthebibliography{#1}%
	\setlength{\parskip}{\bibitemsep}%
	\setlength{\itemsep}{\bibparskip}%
}

\begin{document}
	%-------------------------------------------------------------------------------
	\date{}
	%don't want date printed

	%\title{\Large \bf Manipulation of CT Scans using Deep Learning}
	%\title{\Large \bf Manipulation of 3D CT Scans using Deep Learning}
	%\title{\Large \bf Malicious Tampering of CT Scans using Deep Learning}
	
	%\title{\Large \bf CT-GAN: Malicious Tampering of 3D Medical Imagery using Deep Learning}
	\title{\Large \bf CT-GAN: Malicious Tampering of 3D Medical Imagery using Deep Learning}
	
	%%for single author (just remove % characters)
	\author{
		{\rm Yisroel Mirsky\textsuperscript{1}, Tom Mahler\textsuperscript{1}, Ilan Shelef\textsuperscript{2}, and Yuval Elovici\textsuperscript{1}}\\
		\textsuperscript{1}Department of Information Systems Engineering, Ben-Gurion University, Israel\\
		\textsuperscript{2}Soroka University Medical Center, Beer-Sheva, Israel\\
		yisroel@post.bgu.ac.il, mahlert@post.bgu.ac.il, shelef@bgu.ac.il, and elovici@bgu.ac.il\\\\
		\normalsize\textit{Published in the 28th USENIX Security Symposium (USENIX Security 2019)}\\\\
		\normalsize\textbf{Demo video with pen-test:} \url{https://youtu.be/_mkRAArj-x0}\\
		\normalsize\textbf{Source code and datasets:} \url{https://github.com/ymirsky/CT-GAN}
		\vspace{-.6cm}
	}
	%Second Institution
	%% copy the following lines to add more authors
	%% \and
	%% {\rm Name}\\
	%%Name Institution
	%} % end author
	
	\maketitle
	\thispagestyle{empty}
	%\vspace{-5cm}
	%-------------------------------------------------------------------------------
	\begin{abstract}
		%-------------------------------------------------------------------------------
		%The security of health-care organizations has been lagging behind modern standards. 
		In 2018, clinics and hospitals were hit with numerous attacks leading to significant data breaches and interruptions in medical services. An attacker with access to medical records can do much more than hold the data for ransom or sell it on the black market. 
		
		In this paper, we show how an attacker can use deep-learning to add or remove evidence of medical conditions from volumetric (3D) medical scans. An attacker may perform this act in order to stop a political candidate, sabotage research, commit insurance fraud, perform an act of terrorism, or even commit murder. We implement the attack using a 3D conditional GAN and show how the framework (CT-GAN) can be automated. Although the body is complex and 3D medical scans are very large, CT-GAN achieves realistic results which can be executed in milliseconds.
		
		To evaluate the attack, we focused on injecting and removing lung cancer from CT scans. We show how three expert radiologists and a state-of-the-art deep learning AI are highly susceptible to the attack. We also explore the attack surface of a modern radiology network and demonstrate one attack vector: we intercepted and manipulated CT scans in an active hospital network with a covert penetration test.

	\end{abstract}
	\thispagestyle{empty}
	%keywords: fraud
	
	%-------------------------------------------------------------------------------
	\section{Introduction}
	%-------------------------------------------------------------------------------
	
	% what is CT scans
	Medical imaging is the non-invasive process of producing internal visuals of a body for the purpose of medical examination, analysis, and treatment. In some cases, volumetric (3D) scans are required to diagnose certain conditions. The two most common techniques for producing detailed 3D medical imagery are Magnetic Resonance Imaging (MRI), and CT (Computed Tomography). Both MRI and CT scanner are essential tools in the medical domain. In 2016, there were approximately 38 million MRI scans and 79 million CT scans performed in the United States \cite{stats}.\footnote{245 CT scans and 118 MRI scans per 1,000 inhabitants.}
	
	MRI and CT scanners are similar in that they both create 3D images by taking many 2D scans of the body over the axial plane (from front to back) along the body. The difference between the two is that MRIs use powerful magnetic fields and CTs use X-Rays. As a result, the two modalities capture body tissues differently: MRIs are used to diagnose issues with bone, joint, ligament, cartilage, and herniated discs. CTs are used to diagnose cancer, heart disease, appendicitis, musculoskeletal disorders, trauma, and infectious diseases \cite{john2008ct}.

	% The network which houses CT scans
	Today, CT and MRI scanners are managed though a picture archiving and communication system (PACS). A PACS is essentially an Ethernet-based network involving a central server which (1) receives scans from connected imaging devices, (2) stores the scans in a database for later retrieval, and (3) retrieves the scans for radiologists to analyze and annotate. The digital medical scans are sent and stored using the standardized DICOM format.\footnote{\url{https://www.dicomstandard.org/about/}}
	
	\subsection{The Vulnerability}
	The security of health-care systems has been lagging behind modern standards \cite{Thebigge8:online, Feelingt39:online, Cybersec8:online, COVENTRY201848}. This is partially because health-care security policies mostly address data privacy (access-control) but not data security (availability/integrity) \cite{jalali2018cybersecurity}. Some PACS are intentionally or accidentally exposed to the Internet via web access solutions. Some example products include Centricity PACS (GE Healthcare), IntelliSpace (Philips), Synapse Mobility (FujiFilm), and PowerServer (RamSoft). A quick search on Shodan.io reveals 1,849 medical image (DICOM) servers and 842 PACS servers exposed to the Internet. Recently, a researcher at McAfee demonstrated how these web portals can be exploited to view and modify a patient's 3D DICOM imagery \cite{McAfeeRe34:online}. PACS which are not directly connected to the Internet are indirectly connected via the facility's internal network \cite{huang2019pacs}. They are also vulnerable to social engineering attacks, physical access, and insiders \cite{insider}.
	
	Therefore, a motivated attacker will likely be able to access a target PACS and the medical imagery within it. Later in section \ref{sec:attackmodel} we will discuss the attack vectors in greater detail.
	
	\subsection{The Threat}
\begin{figure}[t]
		\centering
		\includegraphics[width=\columnwidth]{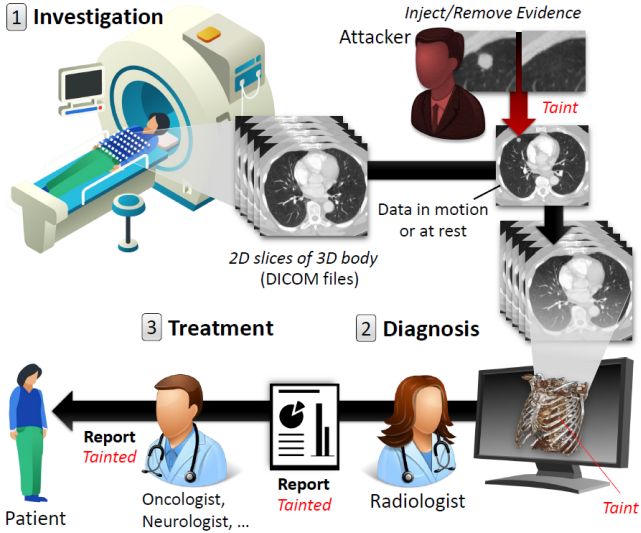}%4a
		\caption{By tampering with the medical imagery between the investigation and diagnosis stages, both the radiologist and the reporting physician believe the fallacy set by the attacker.}
		\label{fig:attack_overview}
		\vspace{-1em}
	\end{figure}

	An attacker with access to medical imagery can alter the contents to cause a misdiagnosis. Concretely, the attacker can add or remove evidence of some medical condition. Fig. \ref{fig:attack_overview} illustrates this process where an attacker injects/removes lung cancer from a scan. 
	
	Volumetric medical scans provide strong evidence of medical conditions. In many cases, a patient may be treated based on this evidence without the need to consider other medical tests. For example, some lesions are obvious or require immediate surgery. Moreover, some lesions will legitimately not show up on other medical tests (e.g., meniscus trauma and some breast cancers). Regardless, even if other tests aren't usually negative, ultimately, the evidence in the scan will be used to diagnose and treat the patient. As a result, an attacker with access to a scan has the power to change the outcome of the patient's diagnosis. For example, an attacker can add or remove evidence of aneurysms, heart disease, blood clots, infections, arthritis, cartilage problems, torn ligaments or tendons, tumors in the brain, heart, or spine, and other cancers.

	There are many reasons why an attacker would want to alter medical imagery. Consider the following scenario: An individual or state adversary wants to affect the outcome of an election. To do so, the attacker adds cancer to a CT scan performed on a political candidate (the appointment/referral can be pre-existing, setup via social engineering, or part of a lung cancer screening program). After learning of the cancer, the candidate steps-down from his or her position. The same scenario can be applied to existing leadership. 
	
	Another scenario to consider is that of ransomware: An attacker seeks out monetary gain by holding the integrity of the medical imagery hostage. The attacker achieves this by altering a few scans and then by demanding payment for revealing which scans have been affected.  
	
	Furthermore, consider the case of insurance fraud: Somebody alters his or her own medical records in order to receive money directly from his or her insurance company, or receive handicap benefits (e.g., lower taxes etc.) In this case, there is no risk of physical injury to others, and the payout can be very large. For example, one can (1) sign up for disability/life insurance, then (2) fake a car accident or other incident, (3) complain of an inability to work, sense, or sleep, then (4) add a small brain hemorrhage or spinal fracture to his or her own scan during an investigation (this evidence is very hard to refute), and then (5) file a claim and receive cash from the insurance company.\footnote{For example, see products such as AIG's Quality of Life insurance.} 
	
	There are many more reasons why an attacker would want to tamper with the imagery. For example: falsifying research evidence, sabotaging another company's research, job theft, terrorism, assassination, and even murder. 
	
	\begin{figure}[t]
		\centering
		\captionof{table}{Summary of an attacker's motivations and goals for injecting/removing evidence in 3D medical imagery.} \label{tab:attack} 
		\includegraphics[width=.95\columnwidth]{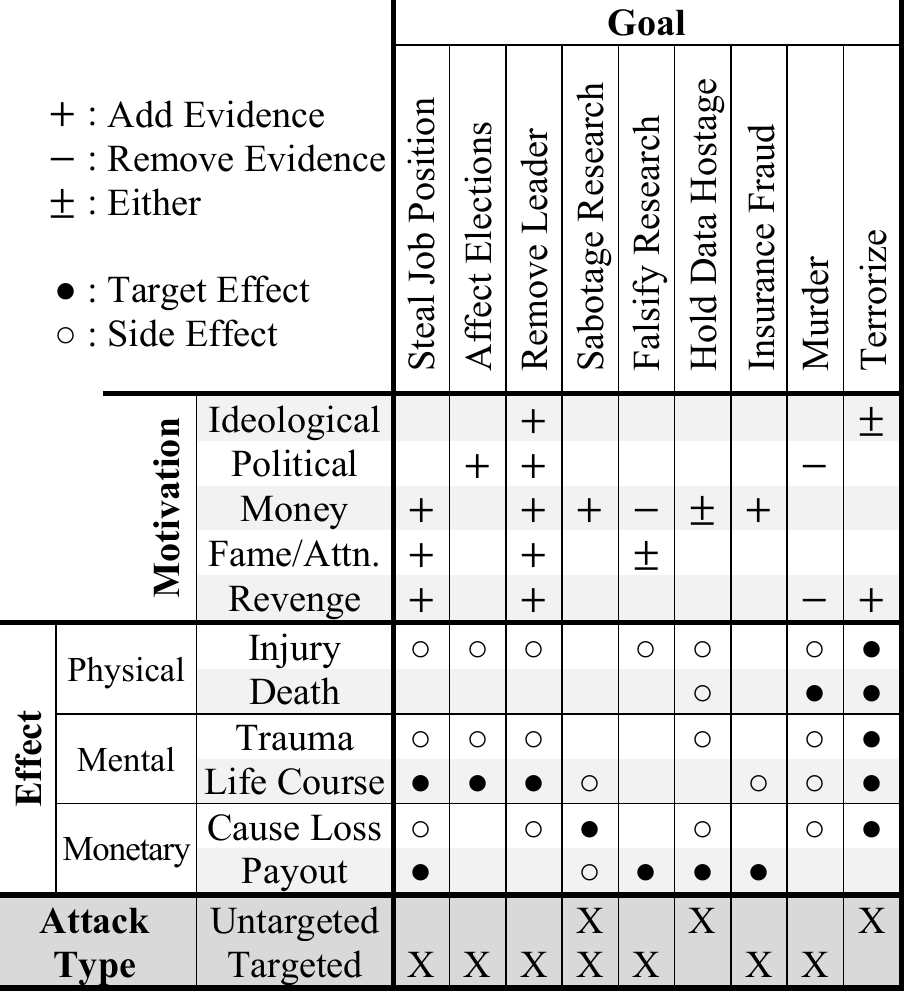}
		\vspace{-1em}
	\end{figure}
	
	Depending on the attacker's goal, the attack may be either untargeted or targeted:
	\begin{description}
		\item[Untargeted Attacks] are where there is no specific target patient. In this case, the attacker targets a victim who is receiving a random voluntary cancer screening, is having an annual scan (e.g., BRACA patients, smokers...), or is being scanned due to an injury. These victims will either have an `incidental finding' when the radiologist reviews the scan (injection) or are indeed sick but the evidence won't show (removal).
		\item[Targeted Attacks] are where there is a specific target patient. In these attacks, the patient may be lured to the hospital for a scan. This can be accomplished by (1) adding an appointment in the system, (2) crafting a cancer screening invite, (3) spoofing the patient's doctor, or (4) tampering/appending the patient's routine lab tests. For example, high-PSA in blood indicates prostate cancer leading to an \textbf{abdominal MRI}, high thyrotropin in blood indicates a brain tumor leading to a \textbf{head MRI}, and metanephrine in urine of hypertensive patients indicates cancer/tumor leading to a \textbf{chest/abdominal CT}
	\end{description}
	
	In this paper we will focus on the injection and removal of lung cancer from CT scans. Table \ref{tab:attack} summarizes attacker's motivations, goals, and effects by doing so. The reason we investigate this attack is because lung cancer is common and has the highest mortality rate \cite{bray2018global}. Therefore, due its impact, an attacker is likely to manipulate lung cancer to achieve his or her goal. We note that the threat, attack, and countermeasures proposed in this paper also apply to MRIs and medical conditions other than those listed above.

	\subsection{The Attack}
	With the help of machine learning, the domain of image generation has advanced significantly over the last ten years \cite{wu2017survey}. In 2014, there was a breakthrough in the domain when Goodfellow et al. \cite{goodfellow2014generative} introduced a special kind of deep neural network called a generative adversarial network (GAN). GANs consist of two neural networks which work against each other: the \textit{generator} and the \textit{discriminator}. The \textit{generator} creates fake samples with the aim of fooling the \textit{discriminator}, and the \textit{discriminator} learns to differentiate between real and fake samples. When applied to images, the result of this game helps the \textit{generator} create fake imagery which are photo realistic. 
	While GANs have been used for positive tasks, researchers have also shown how they can be used for malicious tasks such as malware obfuscation \cite{hu2017generating,rigaki2018bringing} and misinformation (e.g., deepfakes \cite{deepfakes}).

	% However, there are several challenges which make the attack seemingly infeasible:
	%
	%\begin{description}
	%	\item[Problem Complexity] The human anatomy is very complex with many structures and features. Therefore, training a GAN to generate and alter anatomy with the level of accuracy that will fool a human expert is very hard. For example, a radiologist will notice right away if the texture of a tissue is too smooth, if an organ is warped or misshaped, or if a tumor is connected to the body in a way that is impossible. These artifacts may alert the radiologist to investigate the integrity of scan further.
	%	\item[Algorithm Complexity] A 3D CT scan can have over 157 million pixels where the latest advances in GANs can only handle about 2 million pixels (HD images) []. Moreover, processing 3D content is much more complex than 2D imagery. Therefore, training a GAN to alter an entire 3D CT scan will not work out-of-the-box. 
	%	\item[Attack Complexity] Preprocessing a full medical scan for deep learning can take tens of seconds up to a few minutes due to segmentation and interpolation (scaling) [][][]. This delay is problematic for the attacker because a radiologist may analyze a scan immediately after it is taken. Moreover, the attacker may choose to infect the radiologist's viewing application with a malware which alters the content of a targeted scan upon loading it. In these cases, the entire tampering process must be done in a timely manner. 
	%\end{description}
	
	In this paper, we show how an attacker can realistically inject and remove medical conditions with 3D CT scans. The framework, called CT-GAN, uses two conditional GANs (cGAN) to perform in-painting (image completion) \cite{isola2017image} on 3D imagery. For injection, a cGAN is trained on unhealthy samples so that the \textit{generator} will always complete the images accordingly. Conversely, for removal, another cGAN is trained on healthy samples only.
	
	To make the process efficient and the output anatomically realistic, we perform the following steps: (1) locate where the evidence should be inject/removed, (2) cut out a rectangular cuboid from the location, (3) interpolate (scale) the cuboid, (4) modify the cuboid with the cGAN, (5) rescale, and (6) paste it back into the original scan. By dealing with a small portion of the scan, the problem complexity is reduced by focusing the GAN on the relevant area of the body (as opposed to the entire CT). Moreover, the algorithm complexity is reduced by processing fewer inputs\footnote{A 3D CT scan can have over 157 million pixels whereas the latest advances in GANs can only handle about 2 million pixels (HD images).} (pixels) and concepts (anatomical features). This results in fast execution and high anatomical realism. The interpolation step is necessary because the scale of a scan can be different between patients. To compensate for the resulting interpolation blur, we mask the relevant content according to water density in the tissue (Hounsfield units) and hide the smoothness by adding Gaussian white noise. In order to assist the GAN in generating realistic features, histogram equalization is performed on the input samples. We found that this transformation helps the 3D convolutional neural networks in the GAN learn how to generate the subtle features found in the human body. The entire process is automated, meanings that the attack can be deployed in an air gapped PACS.
	
	To verify the threat of this attack, we trained CT-GAN to inject/remove lung cancer and hired three radiologists to diagnose a mix of 70 tampered and 30 authentic CT scans. The radiologists diagnosed 99\% of the injected patients with malign cancer, and 94\% of cancer removed patients as being healthy. After informing the radiologists of the attack, they still misdiagnosed 60\% of those with injections, and 87\% of those with removals.  
	In addition to the radiologists, we also showed how CT-GAN is an effective adversarial machine learning attack. We found that the state-of-the-art lung cancer screening model misdiagnosed 100\% of the tampered patients. Thus, cancer screening tools, used by some radiologists, are also vulnerable to this attack.
	
	This attack is a concern because infiltration of healthcare networks has become common \cite{Thebigge8:online}, and internal network security is often poor \cite{RSAConfe73:online}. Moreover, for injection, the attacker is still likely to succeed even if medical treatment is not performed. This is because many goals rely on simply scaring the patient enough to affect his/her daily/professional life. For example, even if an immediate deletion surgery is not deemed necessary based on the scan and lab results, there will still be weekly/monthly follow-up scans to track the tumor's growth. This will affect the patient's life given the uncertainty of his or her future.

	\subsection{The Contribution}
	
	To the best of our knowledge, it has not been shown how an attacker can maliciously alter the content of a 3D medical image in a realistic and automated way. Therefore, this is the first comprehensive research which exposes, demonstrates, and verifies the threat of an attacker manipulating 3D medical imagery. In summary, the contributions of this paper are as follows:
	\makeatletter
	\renewenvironment{description}%
	{\list{}{\leftmargin=10pt % <------- Adjust this length
			\labelwidth\z@ \itemindent-\leftmargin
			\let\makelabel\descriptionlabel}}%
	{\endlist}
	\makeatother
	\begin{description}
		\item[The Attack Model] We are the first to present how an attacker can infiltrate a PACS network and then use malware to autonomously tamper 3D medical imagery. We also provide a systematic overview of the attack, vulnerabilities, attack vectors, motivations, and attack goals. Finally, we demonstrate one possible attack vector through a penetration test performed on a hospital where we covertly connect a man-in-the-middle device to an actual CT scanner. By performing this pen-test, we provide insights into the security of a modern hospital's internal network.
		\item[Attack Implementation] We are the first to demonstrate how GANs, with the proper preprocessing, can be used to efficiently, realistically, and automatically inject/remove lung cancer into/from large 3D CT scans. We also evaluate how well the algorithm can deceive both humans and machines: radiologists and state-of-the-art AI. We also show how this implementation might be used by an attacker since it can be automated (in the case of an air gapped system) and is fast (in the case of an infected DICOM viewer). 
		\item[Countermeasures] We enumerate various countermeasures which can be used to mitigate the threat. We also provide the reader with best practices and configurations which can be implemented immediately to help prevent this attack.
	\end{description}
	
	For reproducibility and further investigation, we have published our tampered datasets and source code online\footnote{\url{https://github.com/ymirsky/CT-GAN}} along with a pen-test video.\footnote{\url{https://youtu.be/_mkRAArj-x0}}
	
	The remainder of the paper is organized as follows: First we present a short background on GANs. Then, in section \ref{sec:relworks}, we review related works and contrast them ours. In section \ref{sec:attackmodel} we present the attack model and demonstrate one of the attack vectors. In section \ref{sec:attack}, we present CT-GAN's neural architecture, its attack process, and some samples. In section \ref{sec:eval} we evaluate the quality of the manipulations and asses the threat of the attack. Finally, in sections \ref{sec:countermeasures} and \ref{sec:conclusion} we present countermeasures and our conclusion.

	%-------------------------------------------------------------------------------
	\section{Background: GANs}\label{sec:background}
	%-------------------------------------------------------------------------------
	
\begin{figure}[t]
		
		\centering
		\includegraphics[width=\columnwidth]{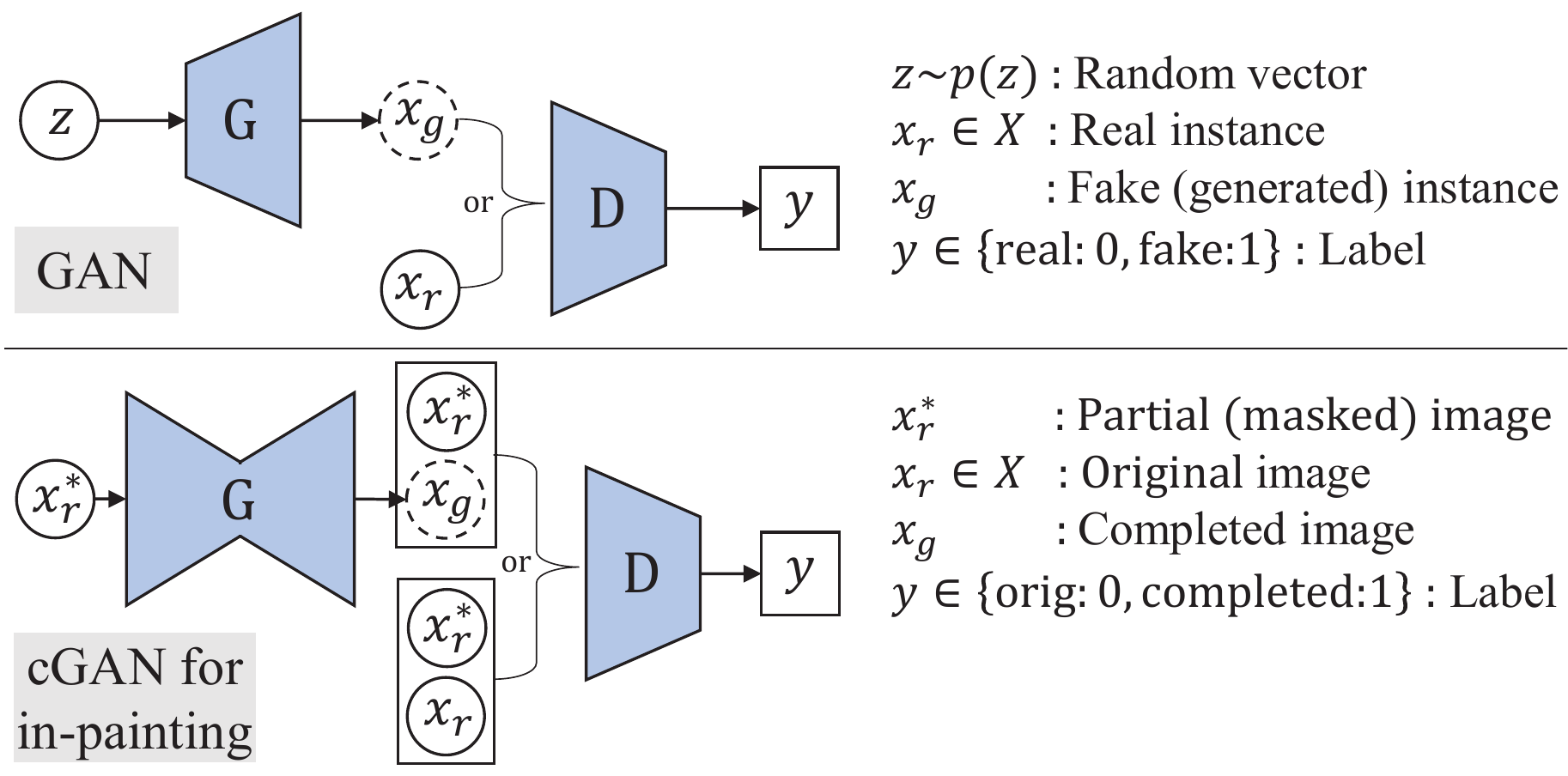}
		\caption{A schematic view of a classic GAN (top) and a cGAN setup for in-painting.}
		\label{fig:gans}
		\vspace{-1.5em}
	\end{figure}

	%move to related works?
	%\subsection{Conditional GANS (cGAN)}
	The most basic GAN consists of two neural networks: the \textit{generator} ($G$) and \textit{discriminator} ($D$). The objective of the GAN is to generate new images which are visually similar to real images found in a sample data distribution $X$ (i.e., a set of images). The input to $G$ is the random noise vector $z$ drawn from the prior distribution $p(z)$ (e.g., a Gaussian distribution). The output of $G$, denoted $x_g$, is an image which is expected to have visual similarity with those in $X$. Let the non-linear function learned by $G$ parametrized by $\theta_g$ be denoted as $x_g = G(z; \theta_g)$. The input to $D$ is either a real image $x_r \in X$ or a generated image $x_g \in G(Z; \theta_g)$. The output of $D$ is the probability that $x_g$ is real or fake. Let the non-linear function learned by $D$ parametrized by $\theta_d$ be denoted as $x_d = D(x; \theta_d)$.
	The top of Fig. \ref{fig:gans} illustrates the configuration of a classic GAN.
	
	%To train a GAN, the following is performed: 
	%
	%\begin{tcolorbox}[breakable,title=\textit{Training Procedure for GANs}]
	%	\small{
	%		\noindent Repeat for $k$ training iterations: 
	%		\begin{enumerate}[leftmargin=*]
	%			\item \textbf{Train $D$}: 
	%			\begin{enumerate}[leftmargin=*]
	%				\item Pull a random batch of samples $x_r \in X$, forward propagate the samples through $D$, compute the error given the label $y=0$, and back propagate the error through $D$ to update $\theta_d$ (using gradient descent or some other variant).
	%				\item Pull a random batch of samples $x_g = G(z; \theta_g)$, forward propagate the samples through $D$, compute the error given the label $y=1$, and back propagate the error through $D$ to update $\theta_d$.
	%				%\item Update $\theta_d$ using the back propagation algorithm on a random batch of samples $x_r \in X$ with the label $y=0$.
	%				%\item Update $\theta_d$ using the back propagation algorithm on a random batch of samples $x_g = G(z; \theta_g)$ with the label $y=1$.
	%			\end{enumerate}
	%			
	%			\item \textbf{Train $G$} using a random batch of vectors $z \sim p(z)$:
	%			\begin{enumerate}[leftmargin=*]
	%				\item Forward propagate $z$ through $G$ and $D$, compute the error at the output of $D$ given the label $y=0$,% (it is set to zero because we want to fool $D$).
	%				back propagate the error through $D$ to $G$ without updating $\theta_d$, and			
	%				continue the back propagation through $G$ while updating $\theta_g$.
	%			\end{enumerate}
	%		\end{enumerate}
	%	}
	%\end{tcolorbox}
	
	It can be seen that $D$ and $G$ are playing a zero-sum game where $G$ is trying to find better (more realistic) samples to fool $D$, while $D$ is learning to catch every fake sample generated by $G$. After training, $D$ is discarded and $G$ is used to generate new samples.
	
	A cGAN is a GAN which has a \textit{generator} and \textit{discriminator} conditioned on an additional input (e.g., class labels). This input extends the latent space $z$ with further information thus assisting the network to generate and discriminate images better. In \cite{isola2017image} the authors propose an image-to-image translation framework using cGANs (a.k.a. pix2pix). There the authors showed how deep convolutional cGANs can be used to translate images from one domain to another. For example, converting casual photos to a Van Gogh paintings. 
	
	One application of the pix2pix framework is in-painting; the process of completing a missing part of an image. When using pix2pix to perform in-painting, the \textit{generator} tries to fill in a missing part of an image based on the surrounding context, and its past experience (other images seen during training). Meanwhile, the \textit{discriminator} tries to differentiate between completed images and original images, given the surrounding context. Concretely, the input to $G$ is a copy of $x_r$ where missing regions of the image are replaced with zeros. We denote this masked input as $x_r^*$. The output of $G$ is the completed image, visually similar to those in $X$. The input to $D$ is either the concatenation $(x_r^*, x_r)$ or $(x_r^*, G(x_r^*; \theta_g))$. The bottom of Fig. \ref{fig:gans} illustrates the described cGAN. The process for training this kind of GAN is as follows:

	\begin{tcolorbox}[breakable,title=\textit{Training Procedure for cGAN In-painting}]
		\small{
			\noindent Repeat for $k$ training iterations: 
			\begin{enumerate}[leftmargin=*]
				\item Pull a random batch of samples $x_r \in X$, and mask the samples with zeros to produce the respective $x_r^*$.
				\item \textbf{Train $D$}: 
				
				\begin{enumerate}[leftmargin=*]
					\item Forward propagate $(x_r^*,x_r)$ through $D$, compute the error given the label $y=0$, and back propagate the error through $D$ to update $\theta_d$.
					\item Forward propagate $(x_r^*,G(x_r^*; \theta_g))$ through $D$, compute the error given the label $y=1$, and back propagate the error through $D$ to update $\theta_d$.
				\end{enumerate}
				
				\item \textbf{Train $G$}:
				\begin{enumerate}[leftmargin=*]
					\item Forward propagate $x_r^*$ through $G$ and then $(x_r^*,G(x_r^*; \theta_g))$ through $D$, compute the error at the output of $D$ given the label $y=0$, % (it is set to zero because we want to fool $D$).
					back propagate the error through $D$ to $G$ without updating $\theta_d$, and			
					continue the back propagation through $G$ while updating $\theta_g$.
				\end{enumerate}
			\end{enumerate}
		}
	\end{tcolorbox}
	Although pix2pix does not use a latent random input $z$, it avoids deterministic outputs by performing random dropouts in the generator during training. this forces the network to learn multiple representations of the data. 
	
	We note that there is a GAN called a CycleGAN \cite{zhu2017unpaired} that can directly translate images between two domains (e.g., benign $\leftrightarrow$ malign). However, we found that the CycleGAN was unable to inject realistic cancer into 3D samples. Therefore, we opted to use the pix2pix model for in-painting because it produced much better results. This may be due to the complexity of the anatomy in the 3D samples and the fact that we had relatively few training samples. Since labeled datasets contain at most a few hundred scans, our approach is more likely to be used by an attacker. Another reason is that in-painting is arguably easier to perform than `style transfer' when considering different bodies. Regardless, in-painting ensures that the final image can be seamlessly pasted back into the scan without border effects.

	%\begin{figure*}[t]
	%	\centering
	%	\includegraphics[width=\textwidth]{GANS3.pdf}
	%	\caption{A schematic view of a classic GAN (top) and a cGAN setup for in-painting.}
	%	\label{fig:gans1}
	%\end{figure*}

	%-------------------------------------------------------------------------------
	\section{Related Work}\label{sec:relworks}
	%-------------------------------------------------------------------------------
	
	The concept of tampering medical imagery, and the use of GANs on medical imagery, is not new. In this section we briefly review these subjects and compare prior results to our work.
	
	\subsection{Tampering with Medical Images}\label{subsec:relworktampering}
	Many works have proposed methods for detecting forgeries in medical images \cite{Singh2017}, but none have focused on the attack itself. The most common methods of image forgery are: copying content from one image to another (image splicing), duplicating content within the same image to cover up or add something (copy-move), and enhancing an image to give it a different feel (image retouching) \cite{Sadeghi2018}. 
	
	Copy-move attacks can be used to cover up evidence or duplicate existing evidence (e.g., a tumor). However, duplicating evidence will raise suspicion because radiologists closely analyze each discovered instance. Image-splicing can be used to copy evidence from one scan to another. However, CT scanners have distinct local noise patterns which are visually noticeable \cite{7018385,duan2017computed}. The copied patterns would not fit the local pattern and thus raise suspicion.
	More importantly, both copy-move and image-splicing techniques are performed using 2D image editing software such as Photoshop. These tools require a digital artist to manually edit the scan. Even if the attacker has a digital artist, it is hard to accurately inject and remove cancer realistically. This is because human bodies are complex and diverse. For example, cancers and tumors are usually attached to nearby anatomy (lung walls, bronchi, etc.) which may be hard to alter accurately under the scrutiny of expert radiologists. Another consideration is that CT scans are 3D and not 2D, which adds to the difficulty. It is also important to note that an attacker will likely need to automate the entire process in a malware since (1) many PACS are not directly connected to the Internet and (2) the diagnosis may occur immediately after the scan is performed.
	
	In contrast to the Photoshopping approach, CT-GAN (1) works on 3D medical imagery, which provide stronger evidence than a 2D scans, (2) realistically alters the contents of a 3D scan while considering nearby anatomy, and (3) can be completely automated. The latter point is important because (1) some PACS are not directly connected to the Internet, (2) diagnosis can happen right after the actual scan, (3) the malware may be inside the radiologist's viewing app.

	%Many works have proposed methods for detecting forgeries \cite{Singh2017}, but none have focused on the attack itself. To evaluate detection algorithms, authors commonly apply transformations to the image but do not actually alter the content or its interpretation. For example, they add noise, rotate the image, apply filters, add contrast, or simply apply JPEG compression. To detect evaluate detection of image splicing [][][] and copy-move attacks [][][], authors make use of Photoshop. 
	
	\subsection{GANs in Medical Imagery}
	
	Since 2016, over 100 papers relating to GANs and medical imaging have been published \cite{yi2018generative}. %GANs can be used in two main ways: using the generator model $G$ to generate new data from existing data or using the discriminator model to detect forged images. 
	These publications mostly relate image reconstruction, denoising, image generation (synthesis), segmentation, detection, classification, and registration.
	We will focus on the use of GANs to generate medical images.
	%For GANs are used for CT image denoising \cite{Yi2018GenerativeReview}: The authors in \cite{Kang2018Cycle} used CycleGAN on multiphase coronary CT angiography 2D images; the authors in \cite{Wolterink2017GenerativeCT} used cGAN for cardiac CT 2D images; the authors in \cite{Yi2018a} used the Pix2Pix framework; and the authors in \cite{You2018} extended the method proposed in \cite{Yang2018Low} to 3D using Pix2Pix on 2D and 3D images.
	%In addition to denoising and synthesis of CT images, GAN are also used for segmentation and classification of CT images. \cite{Yang2017AutomaticNetwork} used an image-to-image network for liver segmentation from 3D CT volumes and evaluated the effect of adversarial loss. 
	
	Due to privacy laws, it is hard to acquire medical scans for training models and students. As a result, the main focus of GANs in this domain has been towards augmenting (expanding) datasets. One approach is to convert imagery from one modality to another. For example, in \cite{Bi2017SynthesisGANs} the authors used cGANs to convert 2D slices of CT images to Positron Emission Tomography (PET) images. In \cite{BenCohen2017VirtualResults, BenCohen2018CrossModalityDetection} the authors demonstrated a similar concept using a fully convolutional network with a cGAN architecture. In \cite{Dou2018UnsupervisedLoss}, the authors converted MRI images to CT images using domain adaptation. In \cite{Jin2018DeepData}, the authors converted MRI to CT images and vice versa using a CycleGAN.
	
	Another approach to augmenting medical datasets is the generation of new instances. In \cite{bermudez2018learning}, the authors use a deep convolutional GAN (DCGAN) to generate 2D brain MRI images with a resolution of 220x172. In \cite{FridAdar2018GANbasedClassification}, the authors used a DCGAN to generate 2D liver lesions with a resolution of 64x64. In \cite{Wolterink2018}, the authors generated 3D blood vessels using a Wasserstien (WGAN). In \cite{baur2018melanogans}, the authors use a Laplace GAN (LAPGAN) to generate skin lesion images with 256x256 resolution. In \cite{madani2018chest}, the authors train two DCGANs for generating 2D chest X-rays (one for malign and the other for benign). However, in \cite{madani2018chest}, the generated samples were down sampled to 128x128 in resolution since this approach could not be scaled to the original resolution of 2000x3000.
	In \cite{Chuquicusma2018} the authors generated 2D images of pulmonary lung nodules (lung cancer) with 56x56 resolution. The author's motivation was to create realistic datasets for doctors to practice on. The samples were generated using a DCGAN and their realism was assessed with help of two radiologists. The authors found that the radiologists were unable to accurately differentiate between real and fake samples.
	
	These works contrast to our work in the following ways:
	\begin{enumerate}
		\item We are the first to introduce the use of GANs as a way to tamper with 3D imagery. The other works focused on synthesizing cancer samples for boosting classifiers, experiments, and training students, but not for malicious attacks. We also provide an overview of how the attack can be accomplished in a modern medical system.
		\item All of the above works either generate small regions of a scan without the context of a surrounding body or generate a full 2D scan with a very low resolution. Samples which are generated without a context cannot be realistically `pasted' back into any arbitrary medical scan. We generate/remove content realistically within existing bodies. Moreover, very low-resolution images of full scans cannot replace existing ones without raising suspicion (especially if the body doesn't match the actual person). Our approach can modify full resolution 3D scans,\footnote{A CT scan can have a resolution from 512x512x600 to 1024x1024x600 and larger.} and the approach can be easily extended to 2D as well.
		\item We are the first to evaluate how well a GAN can fool expert radiologists and state-of-the-art AI in full 3D lung cancer screening. Moreover, in our evaluation, the radiologists and AI were able to consider how the cancer was attached and placed within the surrounding anatomy.
		%\item We are the first work to show how GANs can be used to maliciously and realistically \textit{remove} cancer from 3D CT scans.
	\end{enumerate}

	%-------------------------------------------------------------------------------
	\section{The Attack Model}\label{sec:attackmodel}
	%-------------------------------------------------------------------------------
	
	In this section we explore the attack surface by first presenting the network topology and then by discussing the possible vulnerabilities and attack vectors. We also demonstrate one of the attack vectors on an actual CT scanner.

	\begin{figure*}[t]
		\centering
		\includegraphics[width=\textwidth]{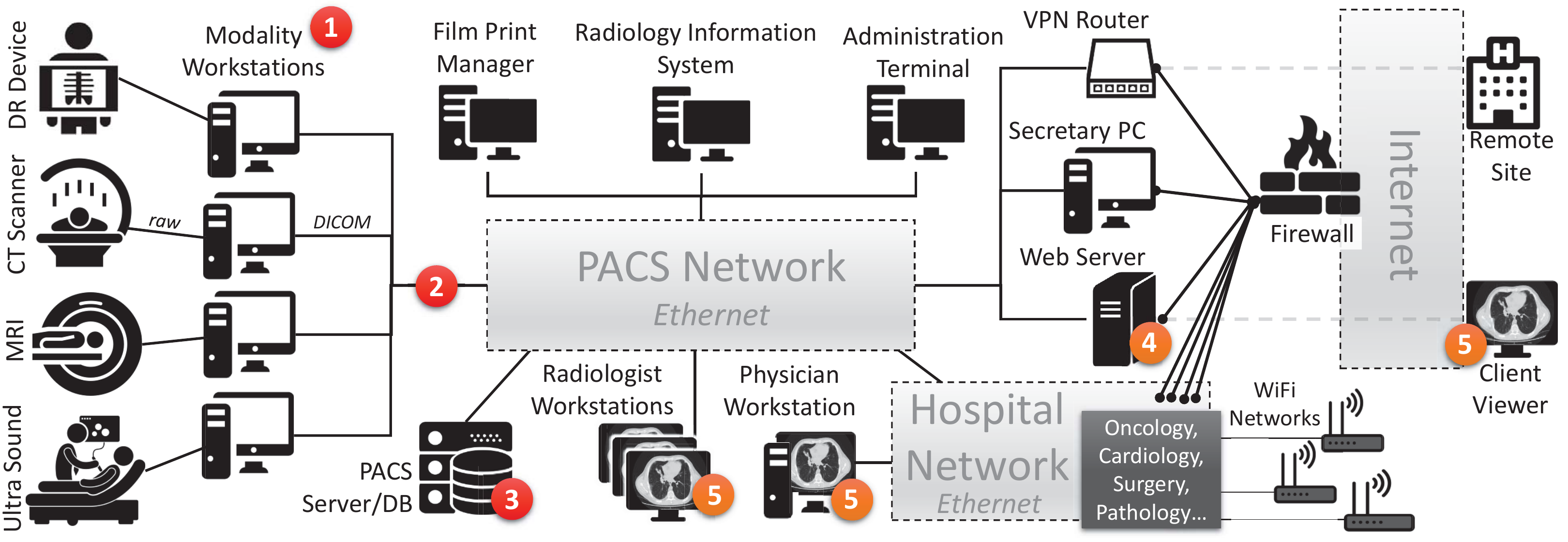}
		\caption{A network overview a PACS in a hospital. 1-3: points where an attacker can tamper with all scans. 4-5: points where an attacker can tamper with a subset of scans.}
		\label{fig:attack_points}
		\vspace{-1em}
	\end{figure*}
	
	\subsection{Network Topology}
	%%note for USENIX version: perhaps less focus on the ordering process, more on the components? static IPs...
	In order to discuss the attack vectors we must first present the PACS network topology. Fig. \ref{fig:attack_points} presents the common network configuration of PACS used in hospitals. The topology is based on PACS literature \cite{hruby2013digital, peck2017clark, carter2018digital, huang2019pacs}, PACS enterprise solutions (e.g., Carestream), and our own surveys conducted on various hospitals. We note that, private medical clinics may have a much simpler topology and are sometimes directly connected to the Internet \cite{McAfeeRe34:online}.

	The basic elements of a PACS are as follows:
	\begin{description}
		\item[PACS Server.] The heart of the PACS system. It is responsible for storing, organizing, and retrieving DICOM imagery and reports commonly via SQL. Although the most facilities use local servers, a few hospitals have transitioned to cloud storage \cite{Cloudbas46:online}. 
		\item[RIS Server.] The radiology information system (RIS) is responsible for managing medical imagery and associated data. Its primary use is for tracking radiology imaging orders and the reports of the radiologists. Doctors in the hospital's internal network can interface with the RIS to order scans, receive the resulting reports, and to obtain the DICOM scans \cite{PictureA43:online}. 
		\item[Modality Workstation.] A PC (typically Windows) which is used to control an imaging modality such as a CT scanner. During an appointment, the attending technician configures and captures the imagery via the workstation. The workstation sends all imagery in DICOM format to the PACS server for storage. 
		\item[Radiologist Workstation.] A radiologist can retrieve and view scans stored on the PACS server from various locations. The most common location is a viewing workstation within the department. Other locations include the radiologist's personal PC (local or remote via VPN), and sometimes a mobile device (via the Internet or within the local network).
		\item[Web Server.] An optional feature which enables radiologists to view of DICOM scans (in the PACS server) over the Internet. The content may be viewed though a web browser (e.g., medDream and Orthanc \cite{jodogne2013orthanc}), an app on a mobile device (e.g., FujiFilm's Synapse Mobility), or accessed via a web API (e.g., Dicoogle \cite{costa2011dicoogle}).
		\item[Administrative Assistant's PC.]  This workstation has both Internet access (e.g., for emails) and access to the PACS network. Access to the PACS is enabled so that the assistant can maintain the devices' schedules: When a patient arrives at the imaging modality, for safety reasons, the technician confirms the patient's identity with the details sent to the modality's workstation (entered by the assistant). This ensures that the scans are not accidentally mixed up between the patients.
		\item[Hospital Network.] Other departments within the hospital usually have access to the PACS network. For example, Oncology, Cardiology, Pathology, and OR/Surgery. In these cases, various workstations around the hospital can load DICOM files from the server given the right credentials. Furthermore, it is common for a hospital to deploy Wi-Fi access points, which are connected to the internal network, for employee access.
	\end{description}

	\subsection{Attack Scenario}
	The attack scenario is as follows: An attacker wants to achieve one of the goals listed in Table \ref{tab:attack} by injecting/removing medical evidence. In order to cause the target effect, the attacker will alter the contents of the target's CT scan(s) before the radiologist performs his or her diagnosis. The attacker will achieve this by either targeting the data-at-rest or data-in-motion. 
	
	Thedata-at-rest refers to the DICOM files stored on the PACS Server, or on the radiologist's personal computer (saved for later viewing). In some cases, DICOM files are stored on DVDs and then transferred to the hospital by the patient or an external doctor. Although the DVD may be swapped by the attacker, it is more likely the interaction will be via the network. The data-in-motion refers to DICOM files being transferred across the network or loaded into volatile memory by an application (e.g., a DICOM viewer). 
	
	We note that this scenario does not apply to the case where the goal is to falsify or sabotage research. Moreover, for insurance fraud, an attacker will have a much easier time targeting a small medical clinic. For simplicity, we will assume that the target PACS is in a hospital.
	
	\subsection{Target Assets}
	
	To capture/modify a medical scan, an attacker must compromise at least one of the assets numbered in Fig. \ref{fig:attack_points}. By compromising one of (1-4), the attacker gains access to every scan. By compromising (5) or (6), the attacker only gains access to a subset of scans. The RIS (3) can give the attacker full control over the PACS server (2), but only if the attacker can obtain the right credentials or exploit the RIS software. The network wiring between the modalities and the PACS server (4) can be used to install a man-in-the-middle device. This device can modify data-in-motion if it is not encrypted (or if the protocol is flawed). 
	
\begin{figure}[b]
		\vspace{-1em}
		\centering
		\includegraphics[width=\columnwidth]{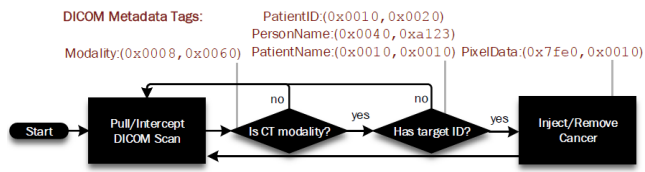}
		\caption{The tampering process of an autonomous malware.}
		\label{fig:malflow}
	\end{figure}

	In all cases, it is likely that the attacker will infect the target asset with a custom malware, outlined in Fig. \ref{fig:malflow}. This is because there may not be a direct route to the PACS via the Internet or because the diagnosis may take place immediately after the scan is taken.
	
	%This malware would automatically 1) locate a DICOM scan with the target user ID stored in its meta-data,\footnote{\raggedright These tags are PatientID:\texttt{(0x0010,0x0020)}, PersonName:\texttt{(0x0040,0xa123)}, and PatientName:\texttt{(0x0010,0x0010)}.}, 2) locate a candidate location to remove/inject the cancer, and then 3) perform the manipulation with a pretrained \textit{generator} algorithm.

	\subsection{Attack Vectors}
	There are many ways in which an attacker can reach the assets marked in Fig. \ref{fig:attack_points}. In general, the attack vectors involve either remote or local infiltration of the facility's network.
	
	\vspace{.3em}
	\noindent\textbf{Remote Infiltration. }
	The attacker may be able to exploit vulnerabilities in elements facing the Internet, providing the attacker with direct access to the PACS from the Internet (e.g., \cite{McAfeeRe34:online}). Another vector is to perform a social engineering attack. For example, a spear phishing attack on the department's administrative assistant to infect his/her workstation with a backdoor, or a phishing attack on the technician to have him install fraudulent updates. 
	
	If the PACS is not directly connected to the Internet, an alternative vector is to (1) infiltrate the hospital's internal network and then (2) perform lateral movement to the PACS. This is possible because PACS is usually connected to the internal network (using static routes and IPs), and the internal network is connected to the Internet (evident from the recent wave of cyber-attacks on medical facilities \cite{Healthca49:online, 11Things18:online,UShospit17:online, Thebigge8:online}). The bridge between the internal network and the PACS is to enable doctors to view scans/reports and to enable the administrative assistant to manage patient referrals \cite{huang2019pacs}. Another vector from the Internet is to compromise a remote site (e.g., a partnered hospital or clinic) which is linked to the hospital's internal network. Furthermore, the attacker may also try to infect a doctor's laptop or phone with a malware which will open a back door into the hospital. 
	
	If the attacker knows that radiologist analyzes scans on his or her personal computer, then the attacker can infect the radiologist's device or DICOM viewer remotely with the malware.
	
	%PHYSICAL ACCESS
	\vspace{.3em}
	\noindent\textbf{Local Infiltration. } The attacker can gain physical access to the premises with a false pretext, such as being a technician from Philips who needs to run a diagnostic on the CT scanner. The attacker may also hire an insider or even be an insider. A recent report shows that $56\%$ of cyber attacks on the healthcare industry come from internal threats \cite{insider}. 
	
	Once inside, the attacker can plant the malware or a back door by (1) connecting a device to exposed network infrastructure (ports, wires, ...) \cite{muniz2015penetration} or (2) by accessing an unlocked workstation. Another vector which does not involve access to a restricted area, is to access to the internal network by hacking Wi-Fi access points. This can be accomplished using existing vulnerabilities such as 'Krack' \cite{vanhoef2017key} or the more recent `BleedingBit' vulnerabilities which have affected many hospitals \cite{Security72:online}.

	%COMPROMISING THE PACS
	\vspace{.3em}
	\noindent\textbf{Compromising the PACS. }
	Once access to the PACS has been achieved, there are numerous ways an attacker can compromise a target asset. Aside from exploiting misconfigurations or default credentials, the attacker can exploit known software vulnerabilities. With regards to PACS servers, some already disclose private information/credentials which can be exploited remotely to create of admin accounts, and have hard-coded credentials.\footnote{\texttt{CVE-2017-14008} and \texttt{CVE-2018-17906}} A quick search on \url{exploit-db.com} reveals seven implemented exploits for PACS servers in 2018 alone. 
	With regards to modality workstations, they too have been found to have significant vulnerabilities \cite{Hospital7:online}. In 2018 the US Department of Homeland Security exposed `low skill' vulnerabilities in Philips' Brilliance CT scanners \cite{Philipsi87:online}. For example, improper authentication, OS command injection, and hard-coded credentials.\footnote{\texttt{CVE-2018-8853}, \texttt{CVE-2018-8857}, and \texttt{CVE-2018-8861}} Other recent vulnerabilities include hard-coded credentials.\footnote{\texttt{CVE-2017-9656}}
	
	Given the state of health-care security, and that systems such as CT scanners are rarely given software updates \cite{Imaginey88:online}, it is likely that these vulnerabilities and many more exist. Once the target asset in the PACS has been compromised, the attacker will be able to install the malware and manipulate the scans of target patients. 
	
	%we found that  it is common to find PACS systems with end-to-end encryption disabled, [**self singed certs?], and default passwords available in administrator interfaces.    
	%Finally, since the PACS is assumed to be air-gapped, many organizations do not enable end-to-end encryption within the PACS nor do they have the medical devices sign the DICOM files with cryptographic signatures. 

	%58 percent of incidents involve insiders compared to just 42 percent tied to external actors \cite{insider}.

	\subsection{Attack Demonstration}
	To demonstrate how an attacker can access and manipulate CT scans, we performed a penetration test on a hospital's radiology department. The pen-test was performed with full permission of the participating hospital. To gain access to all CT scans, we performed a man-in-the-middle attack on the CT scanner using a Raspberry Pi 3B. The Pi was given a USB-to-Ethernet adapter, and was configured as a passive network bridge (without network identifiers). The Pi was also configured as a hidden Wi-Fi access point for backdoor access. We also printed a 3D logo of the CT scanner's manufacturer and glued it to the Pi to make it less conspicuous.
	\begin{figure}[t]
		\centering
		\includegraphics[width=\columnwidth]{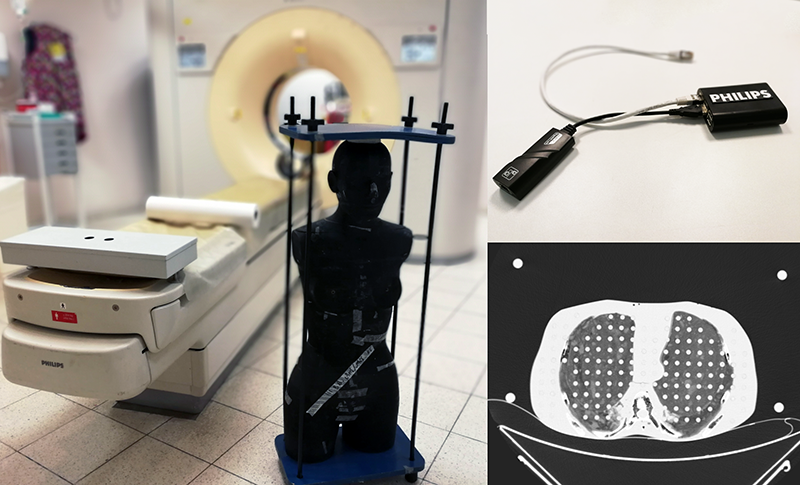}
		\caption{Left: The CT scanner and the medical dummy used to validate the attack. Top-right: the Pi-bridge used to intercept the scans. Bottom-right: one of the dummy's slices, sent by the CT scanner, and intercepted by the Pi-bridge. }
		\label{fig:pentest}
		\vspace{-1em}
	\end{figure}
	The pen-test was performed as follows: First we waited at night until the cleaning staff opened the doors. Then we found the CT scanner's room and installed the Pi-bridge between the scanner's workstation and the PACs network (location \#2 in Fig. \ref{fig:attack_points}). Finally, we hid the Pi-bridge under an access panel in the floor. The entire installation process took 30 seconds to complete. We were able to connect to the Pi wirelessly from the waiting room (approximately 20m away) to monitor the device's status, change the target identifier, etc. 
	
	At this point, an attacker could either intercept scans directly or perform lateral movement through the PACS to other subsystems and install the malware there. To verify that we could intercept and manipulate the scans, we scanned a medical dummy (Fig. \ref{fig:pentest}). We found that the scan of the dummy was sent over the network twice: once in cleartext over TCP to an internal web viewing service, and again to the PACS storage server using TLSv1.2. However, to our surprise, the payload of the TLS transmission was also in cleartext. Moreover, within 10 minutes, we obtained the usernames and passwords of over 27 staff members and doctors due to multicasted Ethernet traffic containing \texttt{HTTP POST} messages sent in cleartext. A video of the pen-test can be found online.\footnote{\url{https://youtu.be/_mkRAArj-x0}}
	
	These vulnerabilities were disclosed to the hospital's IT staff and to their PACS software provider. Though inquiry, we found that it is not common practice for hospitals to encrypt their internal PACs traffic \cite{zetter:online}. One reason is compatibility: hospitals often have old components (scanners, portals, databases, ...) which do not support encryption. Another reason is some PACS are not directly connected to the Internet, and thus is it erroneously thought that there is no need for encryption.

	\begin{figure*}[t]
		\centering
		\includegraphics[width=\textwidth]{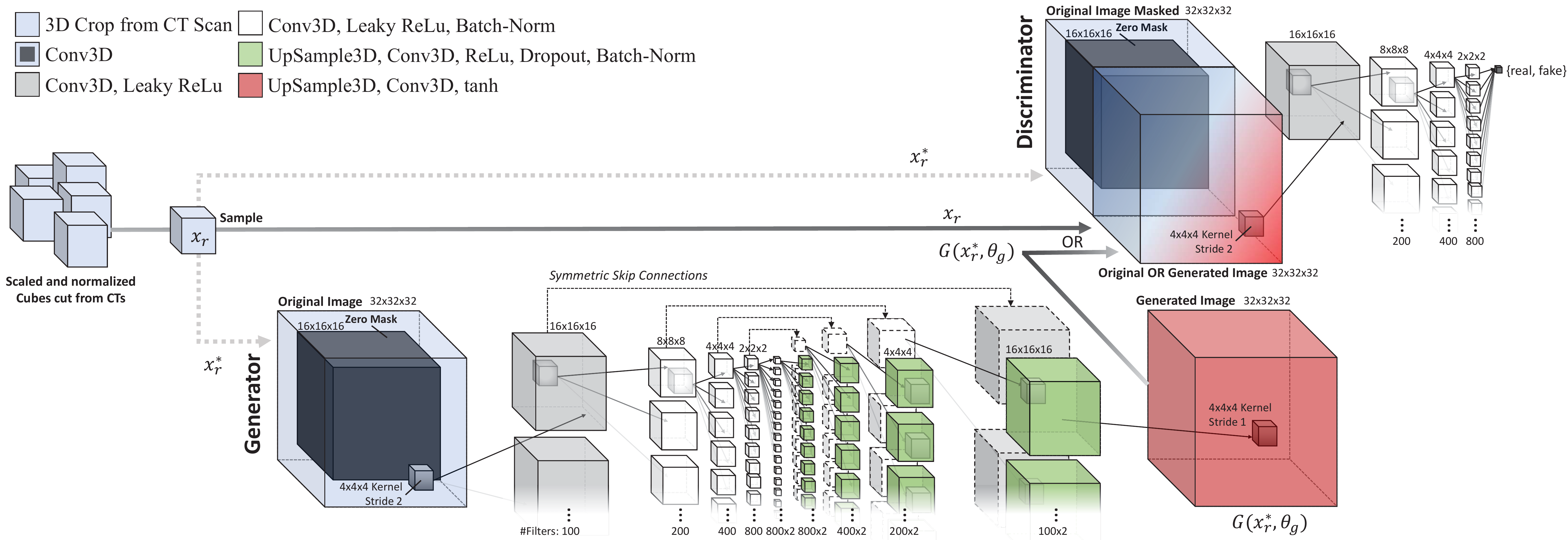}
		\caption{The network architecture, layers, and parameters used for both the injection ($GAN_{\text{inj}}$) and removal ($GAN_{\text{rem}}$) networks.}
		\label{fig:arch}
		\vspace{-1em}
	\end{figure*}
	\begin{figure}[t]
		\centering
		\includegraphics[width=\columnwidth]{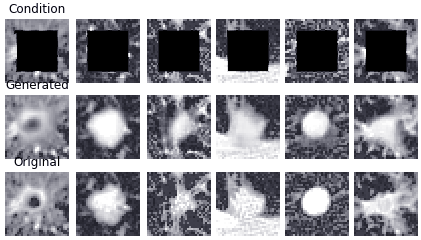}
		\caption{Training samples after 100 epochs showing the middle slice only. Top: the masked sample $x_r^*$ given to both the \textit{generator} $G_{inj}$ and \textit{discriminator} $D_{inj}$. Middle: The in-painted image $x_g$ produced by the $G_{inj}$. Bottom: the ground-truth $x_r$. Note, $D_{inj}$ sees either ($x_r^*$, $x_r$) or ($x_r^*$, $x_g$).}
		\label{fig:train_prog}
		\vspace{-1em}
	\end{figure}

	%-------------------------------------------------------------------------------
	\section{The CT-GAN Framework}\label{sec:attack}
	%-------------------------------------------------------------------------------
	In this section, we present the technique which an attacker can use to add/remove evidence in CT scans. First, we present the CT-GAN architecture and how to train it. Then, we will describe the entire tampering process and present some sample results. As a case study, we will focus on injecting/removing lung cancer.
	
	It is important to note that there are many types of lung cancer. A common type of cancer forms a round mass of tissue called a solitary pulmonary nodule. Most nodules with a diameter less than 8mm are benign. However, nodules which are larger may indicate a malign cancerous growth. Moreover, if \textit{numerous} nodules having a diameter $>8$mm are found, then the patient has an increased risk of primary cancer \cite{macmahon2017guidelines}. For this attack, we will focus on injecting and removing multiple solitary pulmonary nodules.
	
	\subsection{The Neural Architecture}
	A single slice in a CT scan has a resolution of \textit{at least} 512x512 pixels. Each pixel in a slice measures the radiodensity at that location in Hounsfield units (HU). The CT scan of a human's lungs can have over 157 million voxels\footnote{A voxel is the three dimensional equivalent of a pixel.} (512x512x600). In order to train a GAN on an image of this size, we first locate a candidate location (voxel) and then cut out a small region around it (cuboid) for processing. The selected region is slightly larger than needed in order to provide the cGAN with context of the surrounding anatomy. This enables the cGAN to generate/remove lung cancers which connect to the body in a realistic manner. 
	
	To accurately capture the concepts of injection and removal, we use a framework consisting of two cGANs: one for injecting cancer ($GAN_{\text{inj}}$) and one for removing cancer ($GAN_{\text{rem}}$). Both $GAN_{\text{inj}}$ and $GAN_{\text{rem}}$ are deep 3D convolutional cGANs trained to perform in-painting on samples which are $32^3$ voxels in dimension. For the completion mask, we zero-out a $16^3$ cube in the center of the input sample. To inject a large pulmonary nodule into a CT scan, we train $GAN_{\text{inj}}$ on cancer samples which have a diameter of least 10mm. As a result, the trained \textit{generator} completes sample cuboids with similar sized nodules. To remove cancer, $GAN_{\text{rem}}$ is trained using the same architecture, but with samples containing benign lung nodules only (having a diameter $<3$mm). 
	
	The model architecture (layers and configurations) used for both $GAN_{\text{inj}}$ and $GAN_{\text{rem}}$ is illustrated in Fig. \ref{fig:arch}. Overall, $\theta_g$ and $\theta_d$ had $162.6$ million and $26.9$ million trainable parameters respectively ($189.5$ million in total). 
	
	We note that follow up CT scans are usually ordered when a large nodule is found. This is because nodule growth is a strong indicator of cancer \cite{macmahon2017guidelines}. We found that an attacker is able to simulate this growth by conditioning each cancerous training sample on the nodule's diameter. However, the objective of this paper is to show how GANS can add/remove evidence realistically. Therefore, for the sake of simplicity, we have omitted this `feature' from the above model.
	
	\begin{figure*}[t]
		\centering
		\includegraphics[width=\textwidth]{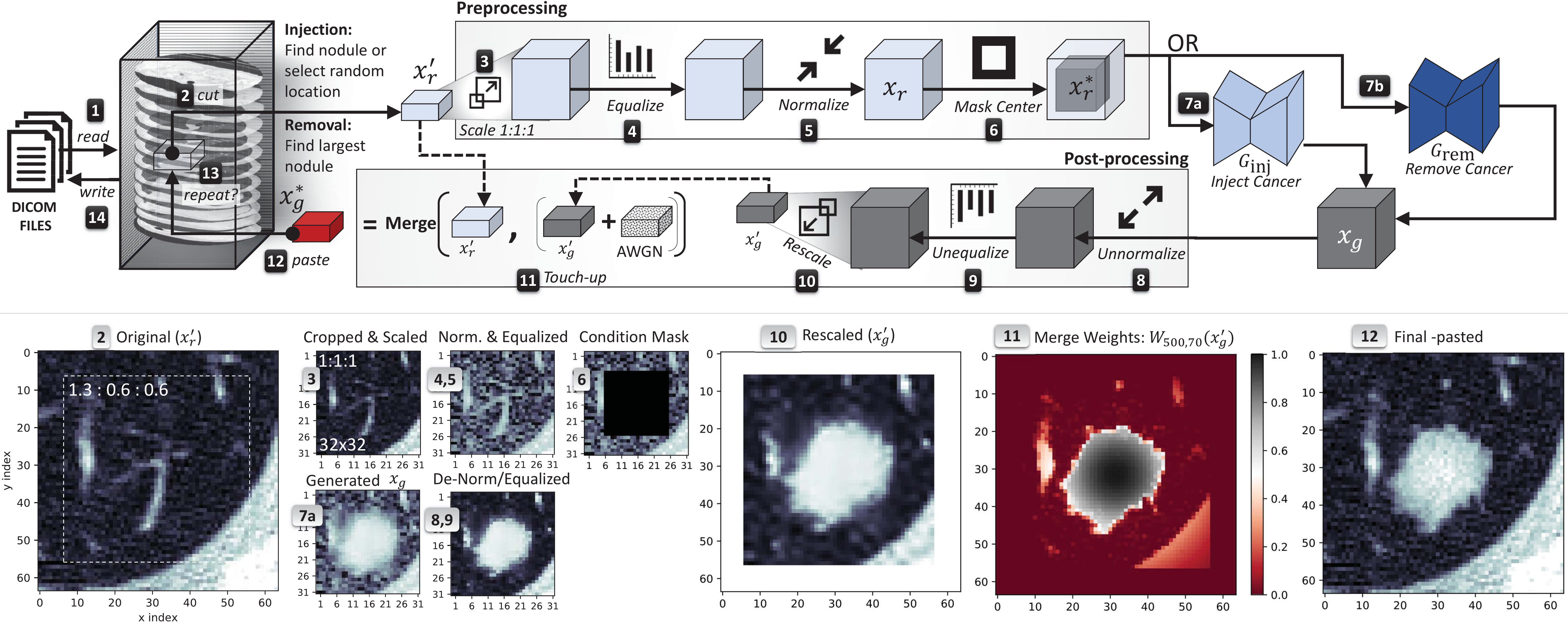}
		\caption{Top: the complete cancer injection/removal process. Bottom: sample images from the \textit{injection} process. The grey numbers indicate from which step the image was taken. The sample 2D images are the middle slice of the respective 3D cuboid.}
		\label{fig:attack_flow}
		\vspace{-1em}
	\end{figure*}

	%what is a conditional GAN
	%our Architecture
	\subsection{Training CT-GAN}
	To train the GANs, we used a free dataset of 888 CT scans collected in the LIDC-IDRI lung cancer screening trial \cite{armato2011lung}. The dataset came with annotations from radiologists: the locations and diameters of pulmonary nodules having diameters greater than $3$mm. In total there were 1186 nodules listed in the annotations.
	
	To create the training set for $GAN_{\text{inj}}$, we extracted from the CT scans all nodules with a diameter between 10mm and 16mm (169 in total). To increase the number of training samples, we performed data augmentation: For each of the 169 cuboid samples, we (1) flipped the cuboid on the $x$, $y$, and $xy$ planes, (2) shifted the cuboid by 4 pixels in each direction on the xy plane, and (3) rotated the cuboid 360 degrees at 6 degree intervals. This produced an additional 66 instances for each sample. The final training set had 11,323 training samples.
	
	To create the training set for $GAN_{\text{rem}}$, we first selected clean CT scans which had no nodules detected by the radiologists. On these scans, we used the nodule detection algorithm from \cite{murphy2009large} (also provided in the dataset's annotations) to find benign micro nodules. Of the detected micro nodules, we selected 867 nodules at random and performed the same data augmentation as above. The final training set had 58,089 samples.
	
	Prior to training the GANs, all of the samples were preprocessed with scaling, equalization, and normalization (described in the next section in detail). Both of the GANs were trained on their respective datasets for 200 epochs with a batch size of 50 samples. Each GAN took 26 hours to complete its training on an NVIDIA GeForce GTX TITAN X using all of the GPU's memory. Fig. \ref{fig:train_prog} shows how well $GAN_{\text{inj}}$ was able to in-paint cancer patterns after 150 epochs.
	
	\subsection{Execution: The Tampering Process}

	In order to inject/remove lung cancer, pre/post-processing steps are required. The following describes the entire injection/removal process as illustrated in Fig. \ref{fig:attack_flow}:
	%\begin{tcolorbox}[breakable,title=\textit{Cancer Injection/Removal Process}]
	%	\small{
	\begin{enumerate}[leftmargin=*]
		\item \textbf{Capture Data.} The CT scan is captured (as data-at-rest or data-in-motion) in either raw or DICOM format using one of the attack vectors from section \ref{sec:attackmodel}.
		\item \textbf{Localize \& Cut.} A candidate location is selected where cancer will be injected/removed, and then the cuboid $x'_r$ is cut out around it.
		\begin{itemize}[noitemsep,wide=0pt, leftmargin=\dimexpr\labelwidth + 2\labelsep\relax]
			\item[] \textbf{Injection:} An injection location can be selected in one of two ways. The fastest way is to take one of the middle slices of the CT scan and select a random location near the middle of the left or right half (see Fig. \ref{fig:average} in the appendix). With $888$ CT scans, this strategy gave us a $99.1$\% successes rate. A more precise way is to execute an existing nodule detection algorithm to find a random micro nodule. To improve speed, the algorithm can be given only a few slices and implemented with early stopping. In our evaluation, we used the algorithm in \cite{murphy2009large}, though many other options are available. 
			\item[] \textbf{Removal:} A removal location can be found by selecting the largest nodule with \cite{murphy2009large}, or by using a pre-trained deep learning cancer detection model.\footnote{Pre-trained models are available here:\\ \url{https://concepttoclinic.drivendata.org/algorithms}}
		\end{itemize}
		\item \textbf{Scale.} $x'_r$ is scaled to the original 1:1:1 ratio using 3D spline interpolation.\footnote{In Python: \texttt{scipy.ndimage.interpolation.zoom}} The ratio information is available in the DICOM meta data with the tags (\texttt{0x0028},\texttt{0x0030}) and (\texttt{0x0018},\texttt{0x0050}). Scaling is necessary because each scan is stored with a different aspect ratio, and a GAN needs consistent units to produce accurate results. To minimize the computations, the cuboid cut in step 2 is cut with the exact dimensions so that the result of the rescaling process produces a $32^3$ cube.		
		\item[4-5.] \textbf{Equalize \& Normalize.} Histogram equalization is applied to the cube to increase contrast. This is a critical step since it enables the GAN to learn subtle features in the anatomy
		% (see \ref{fig:equalization} in the appendix for a visual example). 
		Normalization is then applied using the formula $X_n=2\frac{X-\text{min}(X)}{\text{max}(X)-\text{min}(X)}-1$. This normalization ensures that all values fall on the range of $[-1,1]$ which helps the GAN learn the features better. The output of this process is the cube $x_r$.
		\item[6.] \textbf{Mask.} In the center of $x_r$, a $16^3$ cube is masked with zeros to form $x_r^*$. The masked area will be in-painted (completed) by the \textit{generator}, and the unmasked area is the context). 
		\item[7.] \textbf{Inject/Remove.}  $x_r^*$ is passed through the chosen \textit{discriminator} ($G_{\text{inj}}$ or $G_{\text{rem}}$) creating a new sample ($x_g$) with new 3D generated content.
		\item[8-10.] \textbf{Reverse Preprocessing.} $x_g$ is unnormalized, unequalized, and then rescaled with spline interpolation back to its original proportions, forming $x'_g$. 
		\item[11.] \textbf{Touch-up.} The result of the interpolation usually blurs the imagery. In order to hide this artifact from the radiologists, we added Gaussian noise to the sample: we set $\mu=0$ and $\sigma$ to the sampled standard deviation of $x'_r$. To get a clean sample of the noise, we only measured voxels with values less than $-600$ HU. Moreover, to copy the relevant content into the scan, we merged the original cuboid ($x'_r$) with the generated one ($x'_g$) using a sigmoid weighted average. 
		
		Let $W$ be the weight function defined as
		\begin{equation}\label{eq:alphachannel}
		W_{\alpha,\beta}(x) = \frac{1}{1+e^{-(x+\alpha)/\beta}} * G(x)
		\end{equation}
		where parameter $\alpha$ is the HU threshold between wanted and unwanted tissue densities, and parameter $\beta$ controls the smoothness of the cut edges. The function $G(x)$ returns a 0-1 normalized Gaussian kernel with the dimensions of $x$. $G(x)$ is used to decay the contribution of each voxel the further it is the cuboid's center.  
		
		With $W$, we define the merging function as
		\begin{equation}
		merge_{\alpha,\beta}(x,y) = W_{\alpha,\beta}(x)*x + \left(1-W_{\alpha,\beta}(x)\right)*y
		\end{equation}
		where $x$ is source ($x'_g$) and $y$ is the destination ($x'_r$).
		We found that setting $\alpha=500$ and $\beta=70$ worked best. By applying these touch-ups, the final cuboid $x_g^*$ is produced.
		
		\item[12.] \textbf{Paste.} The cuboid $x_g^*$ is pasted back into the CT scan at the selected location. See Fig. \ref{fig:average} in the appendix for one slice of a complete scan. 
		\item[13.] \textbf{Repeat.} If the attacker is removing cancer, then return to step 2 until no more nodules with a diameter $>3$mm are found. If the attacker is injecting cancer, then (optionally) return to step 2 until four injections have been performed. The reason for this is because the risk of a patient being diagnosed with cancer is statistically greater in the presence of exactly four solitary pulmonary nodules having a diameter $>8$mm \cite{macmahon2017guidelines}.
		\item[14.] \textbf{Return Data.} The scan is converted back into the original format (e.g. DICOM) and returned back to the source.
	\end{enumerate}
	%	}
	%\end{tcolorbox}

\begin{figure}[t]
		\centering
		\includegraphics[width=.48\columnwidth]{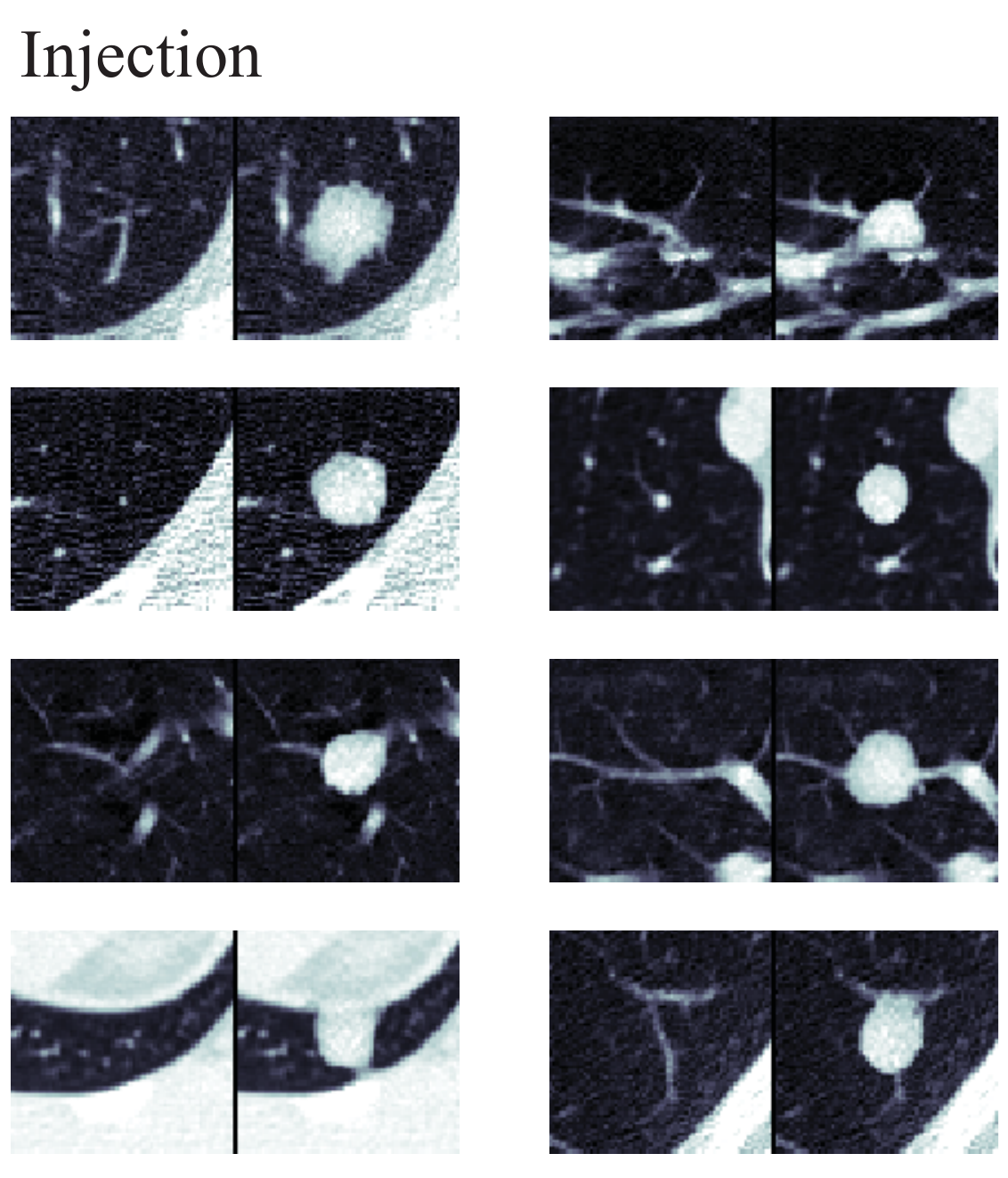} \hspace{.5em}
		\includegraphics[width=.48\columnwidth]{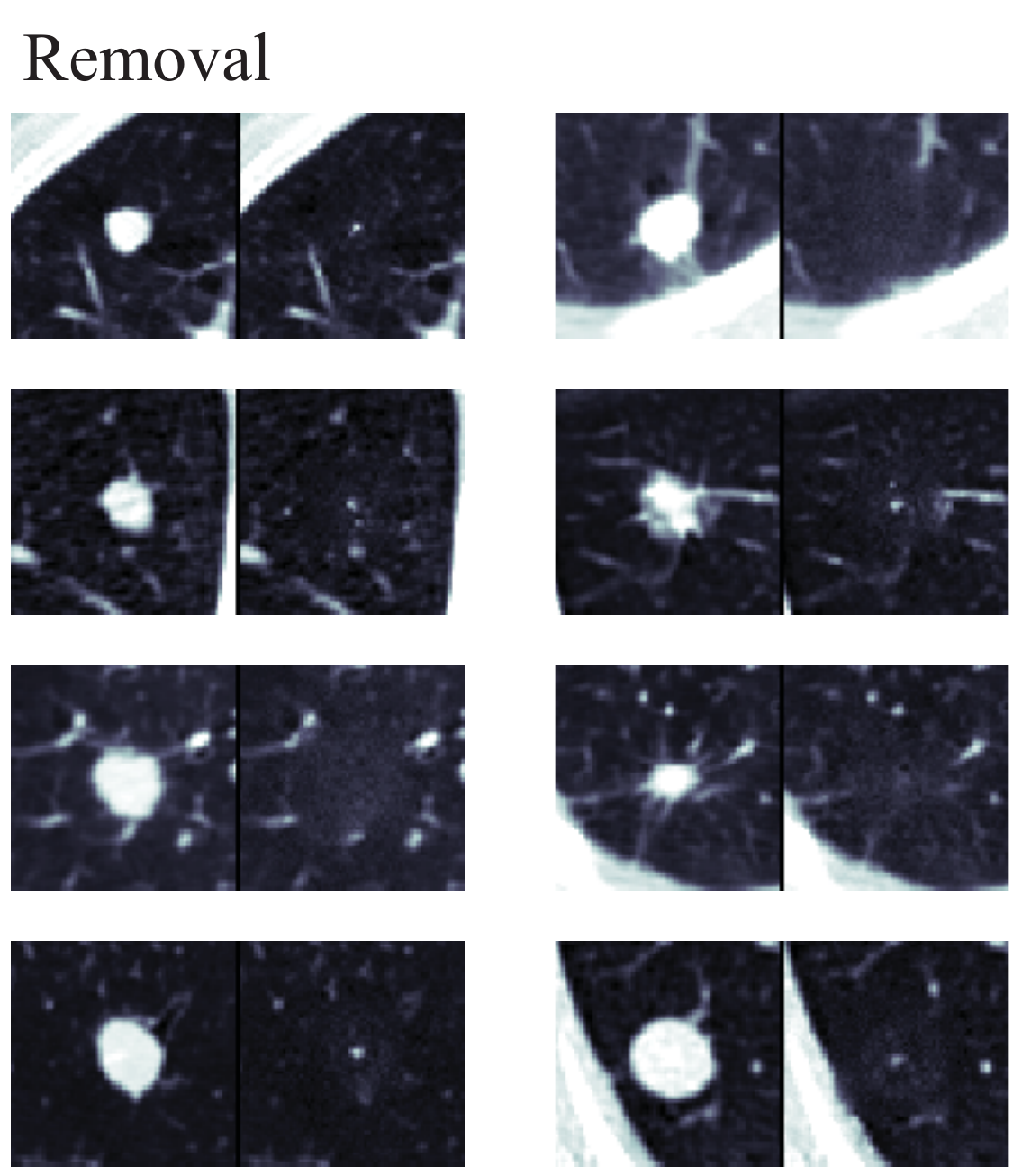}
		
		\caption{Sample injections (left) and removals (right). For each image, the left side is before tampering and the right side is after. Note that only the middle 2D slice is shown.}
		\label{fig:various_examples}
		\vspace{1em}
		\centering
		\includegraphics[width=.48\columnwidth]{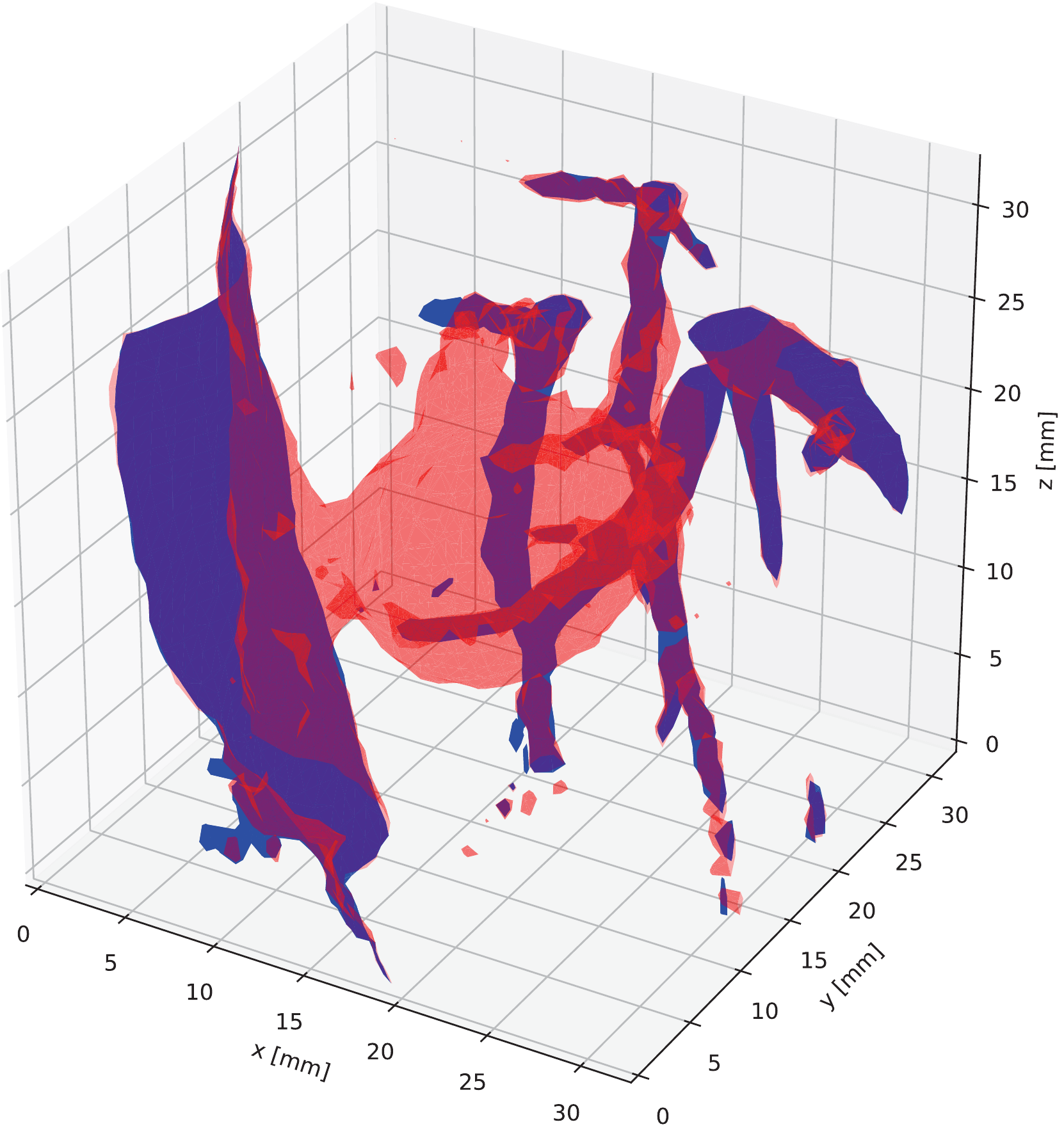}
		\includegraphics[width=.48\columnwidth]{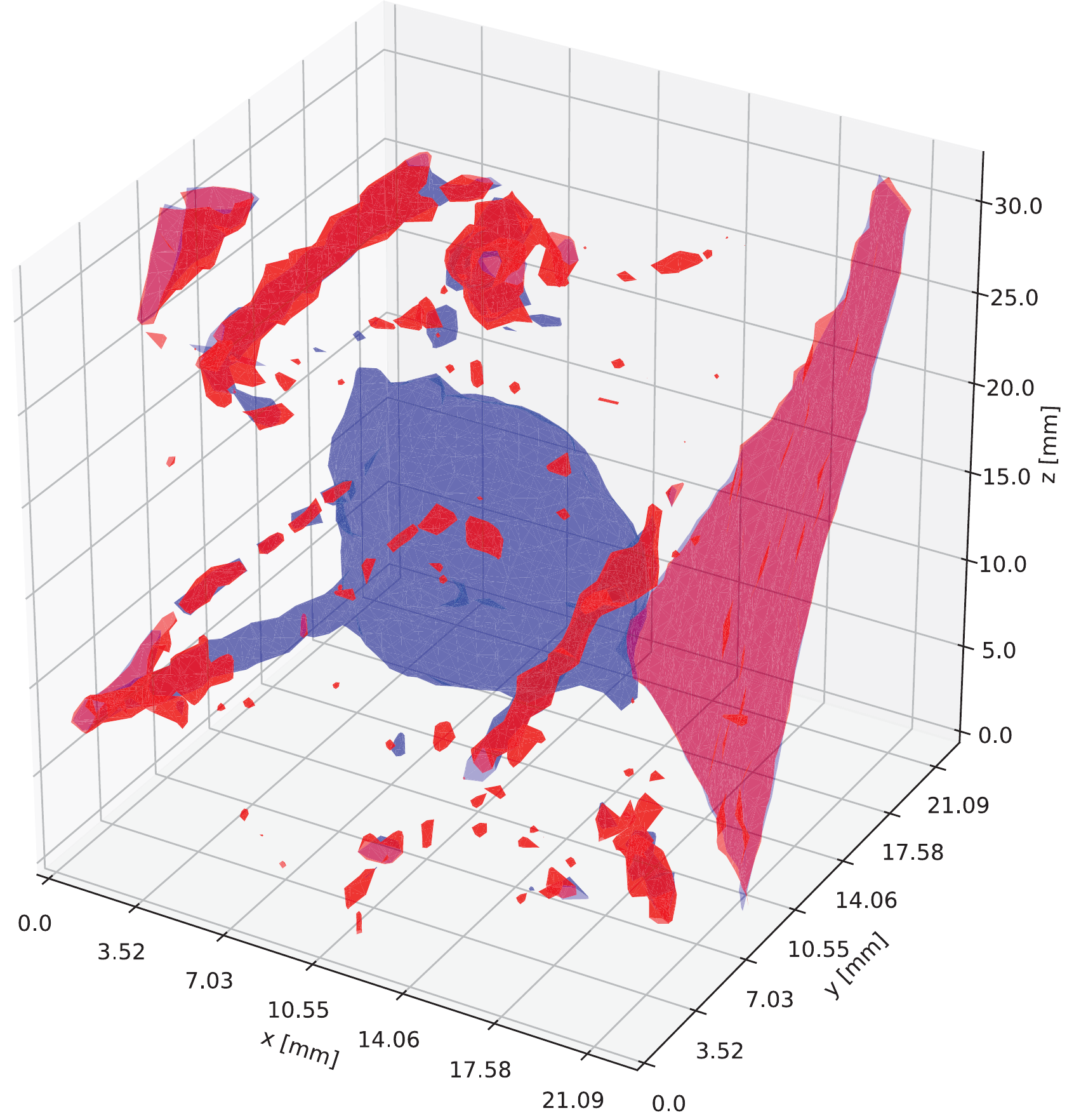}
		\caption{A 3D sample of injection (left) and removal (right) before (blue) and after (red) tampering with the CT scan.}
		\label{fig:3dexample}
	\end{figure}

	The quality of the injection/removal process can be viewed in figures \ref{fig:various_examples} and \ref{fig:3dexample}. Fig. \ref{fig:various_examples} presents a variety of examples before and after tampering, and Fig. \ref{fig:3dexample} provides a 3D visualization of a cancer being injected and removed. More visual samples can be found in the appendix (figures \ref{fig:various_examples_extra} and \ref{fig:inj_slices_extra}). 
	
	We note that although some steps involve image touch-ups, the entire process is automatic (unlike Photoshop) and thus can be deployed in an autonomous malware or inside a viewing application (real-time tampering). We note that the same neural architecture and tampering process works on other modalities and medical conditions. For example, Fig. \ref{fig:braintumors} in the appendix shows CT-GAN successfully injecting brain tumors into MRI head scans.
	
	\begin{figure}
		\centering
		\captionof{table}{Summary of the scans and the relevant notations} \label{tab:dataset} 
		\includegraphics[width=.95\columnwidth]{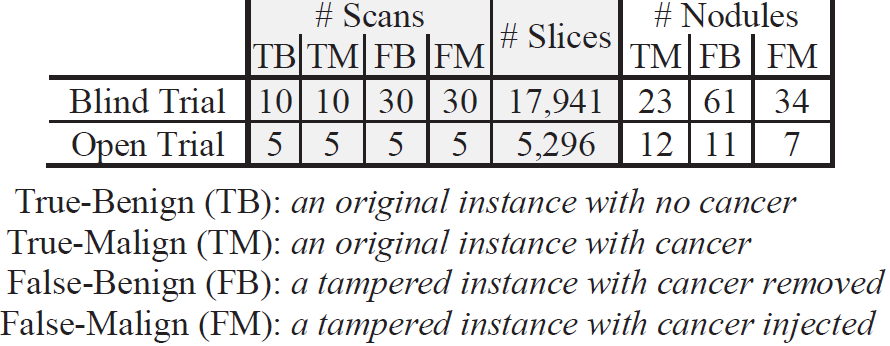}
		\vspace{-1em}
	\end{figure}
	
	\begin{figure*}[t]
		\centering
		\captionof{table}{Cancer Detection Performance - Blind Trial} \label{tab:conf_exp1} 
		\includegraphics[width=\textwidth]{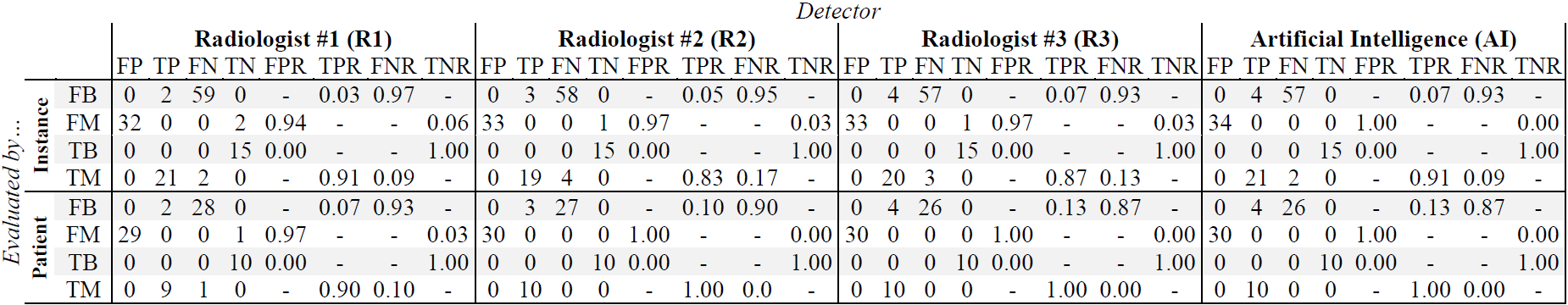}
		\vspace{-2em}
	\end{figure*}

	\begin{figure}[t]
		\centering
		\captionof{table}{Attack Detection Confusion Matrix - Open Trial\\Evalauted by Instance} \label{tab:conf_exp2} 
		\includegraphics[width=.7\columnwidth]{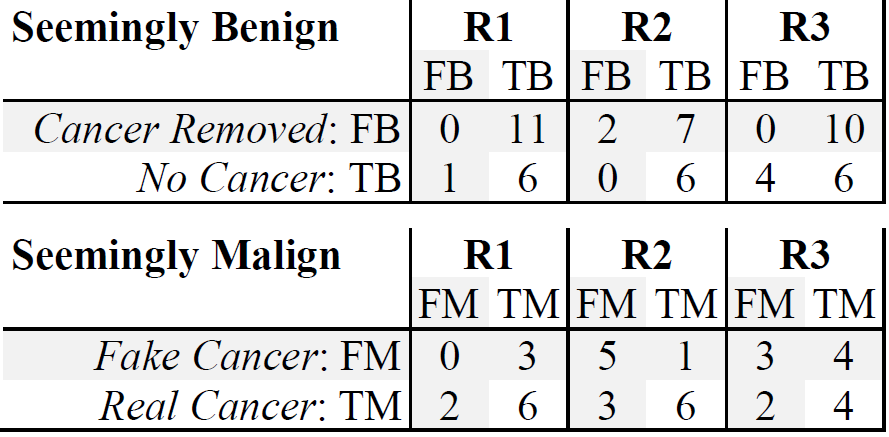}
	\end{figure}
	
	\begin{figure}[t]
		\centering
		\includegraphics[width=\columnwidth]{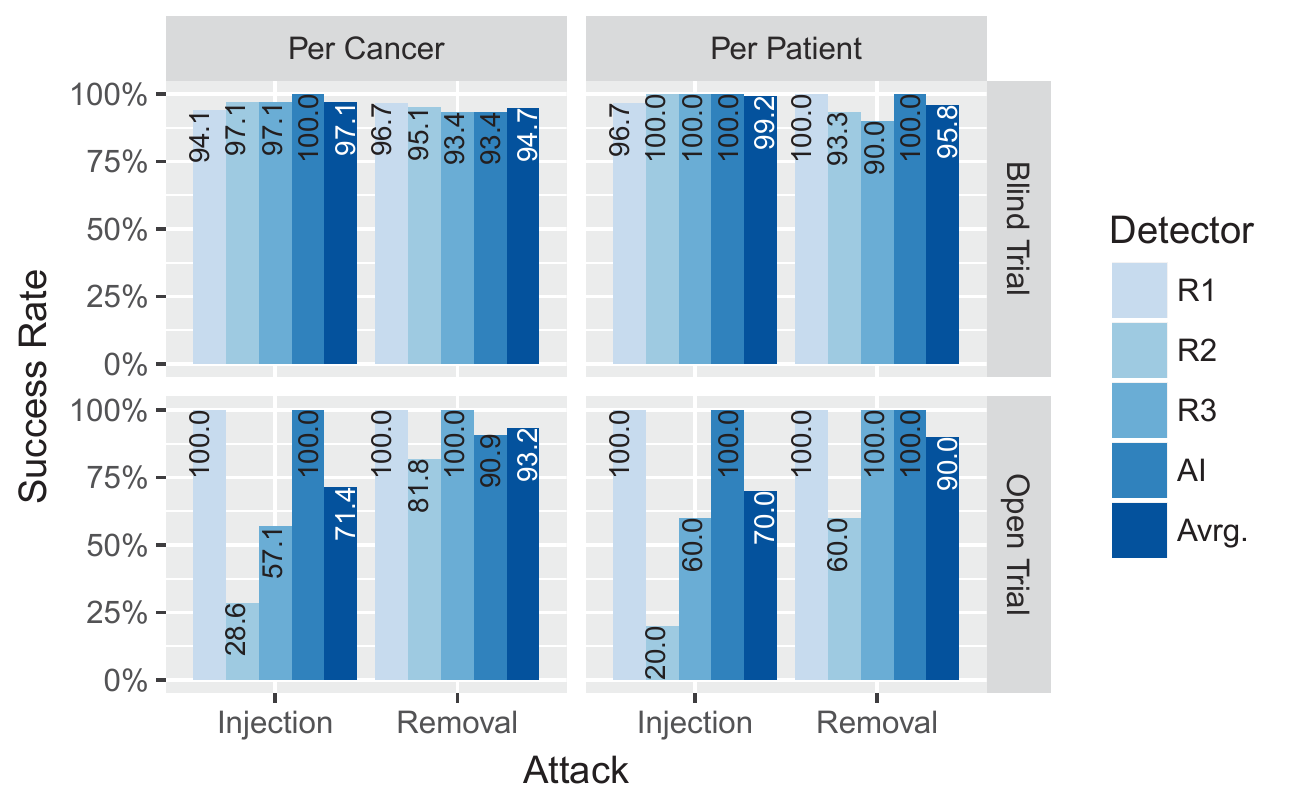}
		\caption{Attack success rates - Both Trials.}
		\label{fig:exp1_succ}
		\vspace{-1em}
	\end{figure}
	\begin{figure}[t]
		\centering
		\includegraphics[width=\columnwidth]{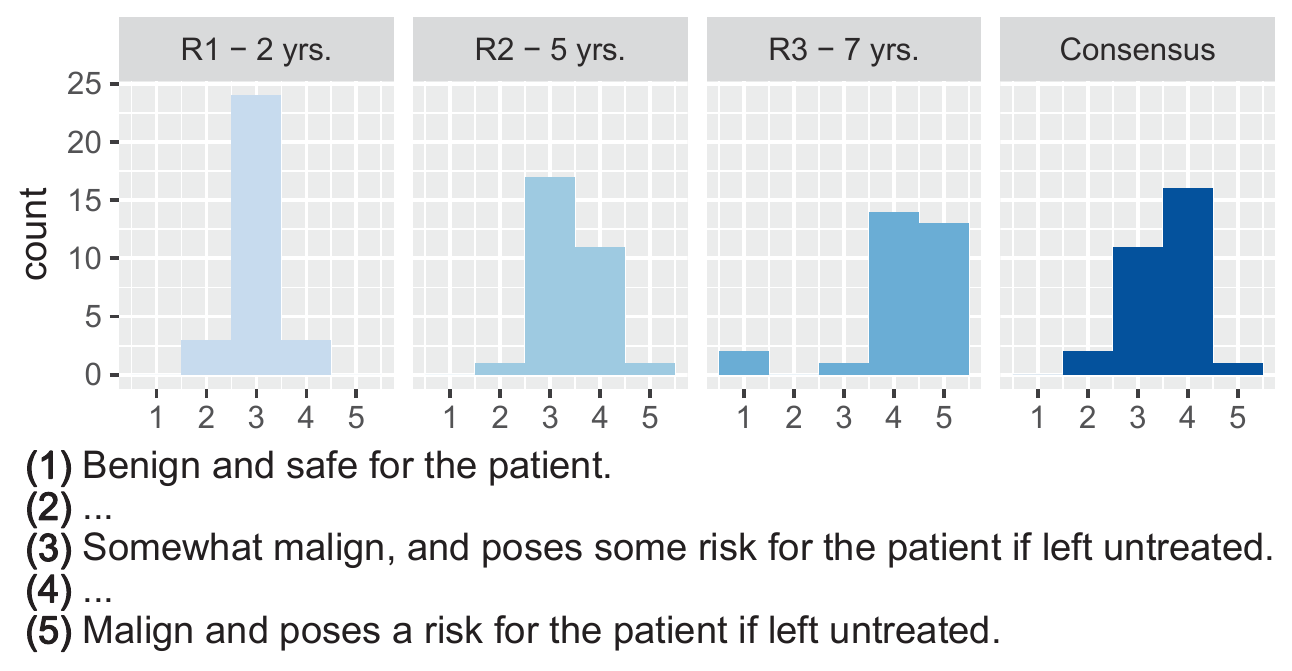}
		\caption{Malignancy of injected cancers (FM) - Blind Trial.}
		\label{fig:exp1_malig}
	\end{figure}
	
	\begin{figure}[t]
		\centering
		\includegraphics[width=\columnwidth]{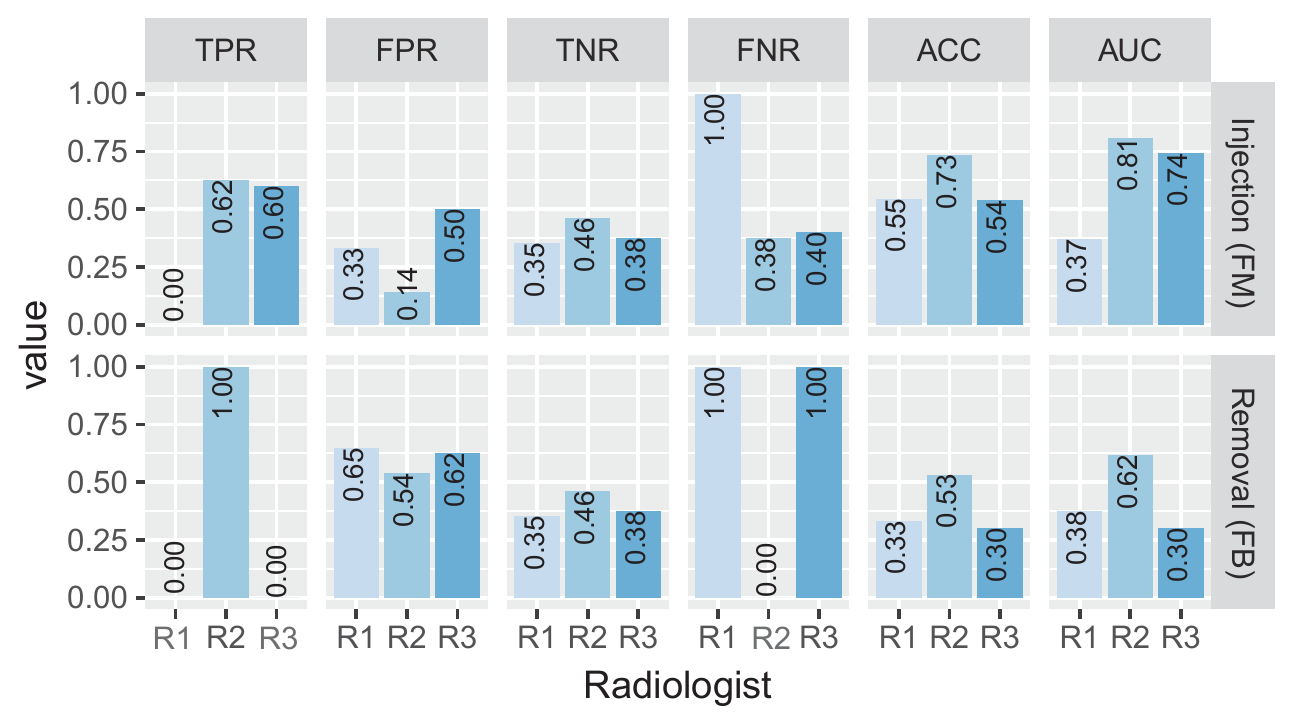}
		\caption{Attack detection performance - Open Trial.}
		\label{fig:exp2_metrics}
		\vspace{-1em}
	\end{figure}
	\begin{figure}[t]
		\centering
		\includegraphics[width=\columnwidth]{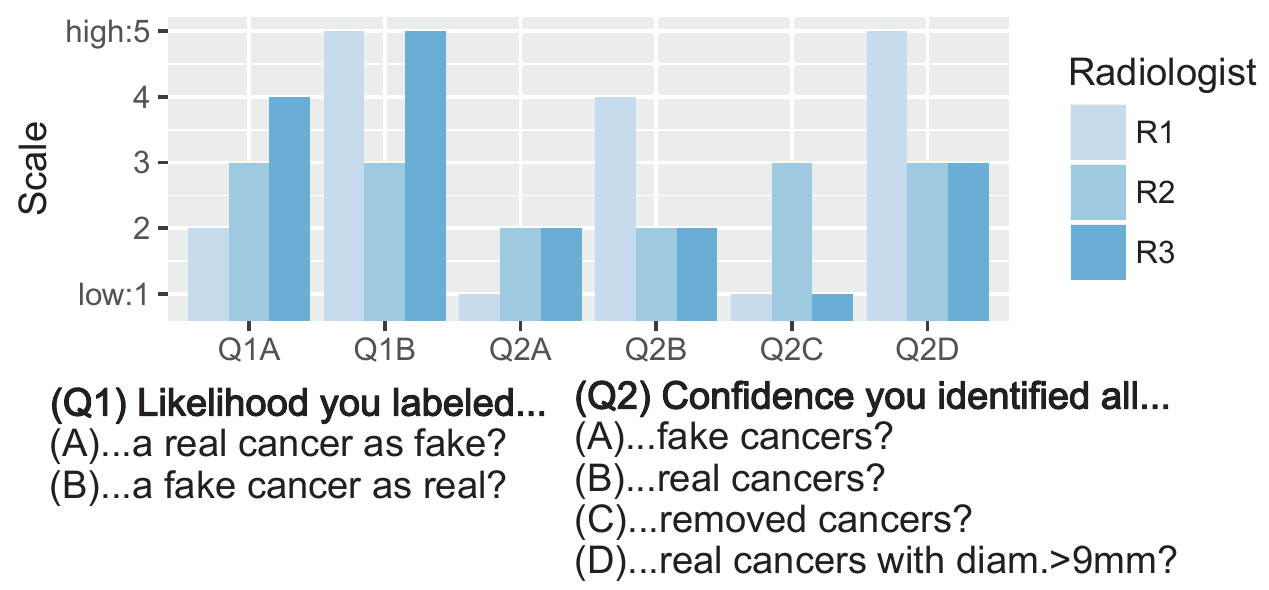}
		\caption{Confidence in detecting attacks - Open Trial.}
		\label{fig:exp2_likerts}
		\vspace{-1em}
	\end{figure}
	
	%-------------------------------------------------------------------------------
	\section{Evaluation}\label{sec:eval}
	%-------------------------------------------------------------------------------
	
	In this section we present our evaluation on how well the CT-GAN attack can fool expert radiologists and state-of-the-art AI.
	
	\subsection{Experiment Setup}
	To evaluate the attack, we recruited three radiologists (denoted \textbf{R1}, \textbf{R2}, and \textbf{R3}) with 2, 5, and 7 years of experience respectively. We also used a trained lung cancer screening model (denoted \textbf{AI}), the same deep learning model which won the 2017 Kaggle Data Science Bowl (a \$1 million competition for diagnosing lung cancer).\footnote{Source code and model available here: \url{https://github.com/lfz/DSB2017}} 
	
	The experiment was performed in two trials: blind and open. In the blind trial, the radiologists were asked to diagnose 80 \textit{complete} CT scans of lungs, but they were not told the purpose of the experiment or that some of the scans were manipulated. In the open trial, the radiologists were told about the attack, and were asked to identify fake, real, and removed nodules in 20 CT scans. In addition, the radiologists were asked to rate the confidence of their decisions. After each trial, we gave the radiologists a questionnaire to assess how susceptible they were to the attacks. In all cases, the radiologists were asked to only detect and diagnose pulmonary nodules which have a diameter greater than 3mm. 
	
	The CT scans were taken from the LIDC-IDRI dataset \cite{armato2011lung}. The set of CT scans used in each trial and the notations used in this section are available in Table \ref{tab:dataset}. 
	
	False benign (FB) and true malign (TM) scans truthfully contained at least one nodule with a diameter between 10mm and 16mm. FB scans were made by removing every nodule in the scan. FM scans were made by randomly injecting 1-4 nodules into a benign scan, where the injected nodules had a diameter of 14.4mm on average. In total, there were 100 CT scans analyzed by each of the radiologists, and the radiologists spent approximately 10 minutes analyzing each of these scans.
	
	We note that the use of three radiologists is common practice in medical research (e.g., \cite{esteva2017dermatologist}). Moreover, we found that radiologists (and AI) significantly agreed with each other's diagnosis per patient and per nodule. We verified this agreement by computing Fliess' kappa \cite{conger1980integration} (a statistical interrater reliability measure) which produced an excellent kappa of 0.84 (p-value $<0.0001$). Therefore, adding more radiologists will likely not affect the results.

	\subsection{Results: Blind Trial}
	In Table \ref{tab:conf_exp1} we present the cancer detection performance of the radiologists and AI. The table lists the number of false-positives (FP - \textit{detected a non-existent cancer}), true-positives (TP - \textit{detected a real cancer}), false-negatives (FN - \textit{missed a real cancer}), and their respective rates. The TCIA annotations (nodule locations) were used as our ground truth for measuring the performance on FB and TM scans. We evaluated these metrics per instance of cancer, and per patient as a whole. All four detectors performed well on the baseline (TB and TM) having an average TPR of 0.975 and a TNR of 1.0 in diagnosing the patients, meaning that we can rely on their diagnosis.
	
	The top of Fig. \ref{fig:exp1_succ} summarizes the attack success rates for the blind trial. In general, the attack had an average success rate of 99.2\% for cancer injection and 95.8\% for cancer removal. The AI was fooled completely which is an important aspect since some radiologists use AI tools to support their analysis (e.g. the Philips IntelliSite Pathology Solution). The radiologists were fooled less so, primarily due to human error (e.g., missing a nodule). When asked, none of the radiologists reported anything abnormal with the scans with the exception of R2 who noted some noise in the area of one removal (FB). This may be attributed to ``inattentional blindness,'' where one may miss an obvious event (artifacts) while engaged in a different task (searching for large nodules). In \cite{drew2013invisible}, the authors showed that this phenomenon also affects radiologists.
	
	With regards to the injected cancers (FM), the consensus among the radiologists was that one-third of the injections require an immediate surgery/biopsy, and that all of the injections require follow-up treatments/referrals. When asked to rate the overall malignancy of the FM patients, the radiologists said that nearly all cases were significantly malign and pose a risk to the patient if left untreated. Fig. \ref{fig:exp1_malig} summarizes radiologists' ratings of the FM patients. One interesting observation is that the malignancy rating increased with the experience of the radiologist. Finally, we note that an attacker could increase the overall malignancy of the injections if CT-GAN were trained only on samples with high malignancy and/or a larger diameter.
	
	\subsection{Results: Open Trial}
	
	In Table \ref{tab:conf_exp2} we present the radiologists' attack detection performance with knowledge of the attack. Fig. \ref{fig:exp2_metrics} summarizes these results and provides the radiologists' accuracy (ACC) and area under the curve (AUC). An AUC of 1.0 indicates a perfect binary classifier, whereas an AUC of 0.5 indicates random guessing. The results show that the radiologists could not consistently tell the difference between real and fake cancers or identify the locations of removed cancers. 
	
	With regards to the attack success rates (bottom of Fig. \ref{fig:exp1_succ}), knowledge of the attack did not significantly affect cancer removal (90\% from 95.8\%). However, the success of the cancer injection was affected (70\% from 99.2\%). Moreover, R2 also picked up on a particular pattern which gave away several instances. This is a promising result, since it indicates that a portion of CT-GAN's attacks can be mitigated by educating radiologists. However, aside from low accuracy (61\% for detecting an injection and 39\% for detecting a removal), there was a significant number of false positives. With a high likelihood of making a mistake, a radiologist may choose not to report abnormalities. This is also apparent from the low confidence scores which the radiologists gave decisions (see Fig. \ref{fig:exp2_likerts}).  
	
	In summary, both radiologists and AI are highly susceptible to CT-GAN's image tampering attacks. Although knowledge of the attack can help mitigate some cases of cancer injection, the error rates and confidence scores suggest that the attack may not be reported in most cases.

	\subsection{Baseline Analysis}
	As mentioned earlier, the attacker cannot use Photoshop on the images. This may be due to a time restriction: the radiologist may perform the diagnosis immediately after the scan has been taken, or the malware may be in the viewing application and operating in real-time. This may also be due to a connectivity restriction: many PACS do not have a direct connection to the Internet. Therefore, the attacker would need to infect the PACS with a malware that can perform the tampering autonomously.
	
	However, for cancer injection, one may argue that there are simpler \textit{automated} methods than using a GAN. For example, the malware could simply paste a pre-saved cropped image of cancer into a scan at a candidate location (found using steps 1 and 2 in Fig. \ref{fig:attack_flow}). Therefore, to validate the use of a GAN, we evaluated this splice attack as a baseline comparison.
	
	The experiment was setup as follows. To create our set of `pre-saved cancers' we carefully cut out ten 32x32x32 cubes of real cancerous nodules from \cite{armato2011lung}. We then removed the irrelevant background tissue by producing transparency (alpha) channels using equation (\ref{eq:alphachannel}). 
	
	The evaluation dataset contained 20 scans in total: 10 FM and 10 TM. To create a FM scan, we took a random healthy scan and used steps 1 and 2 of Fig. \ref{fig:attack_flow} to select a candidate injection location. We then pasted one of the pre-saved cancers into the scan at that location using the transparency channel. To create the FM scans, we randomly selected 10 sick patients with exactly one nodule each.
	
	To evaluate the baseline attack, we asked radiologist \textbf{R2} to analyze the 20 scans and determine which of the scans were authentic (TM) and which were tampered (FM) --similar to the open trial. This task took the radiologist approximately two hours to perform.
	
	In summary, \textbf{R2} correctly labeled all of the tampered scans as fake. The only error was that \textbf{R2} mislabeled three of the authentic scans as fake. These results indicate that the baseline cancer injection attack cannot trick a radiologist, in comparison to CT-GAN which succeeds nearly every time.  
	
\begin{figure}
		\centering
		\includegraphics[width=\columnwidth]{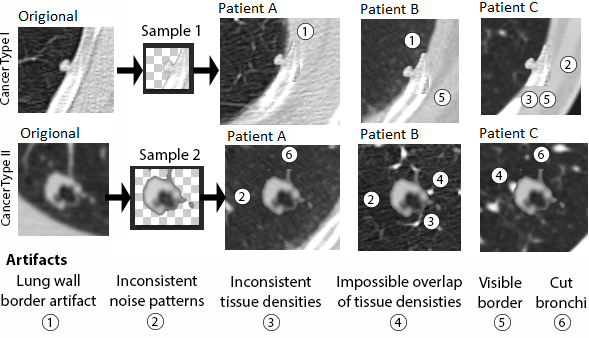}
		\caption{An illustration showing artifacts which can occur when using an \textit{unsupervised} splice attack instead of CT-GAN. Only the middle slice is shown.}
		\label{fig:splice}
		\vspace{-1em}
	\end{figure}

	The reason the baseline attack failed to trick \textbf{R2} is because the process of automatically pasting cancer creates obvious artifacts. This is because the pasted samples ignore the surrounding anatomy and may contain inconsistent noise patterns (textures). Fig. \ref{fig:splice} illustrates some examples of these abnormalities such as cut bronchi, inconsistent noise patterns, and unaligned borders. CT-GAN does not produce these artifacts because it uses in-painting which considers the original content and surrounding anatomy.
	
	%We also note that if an attacker were to simply swap an entire scan with an untampered malign/benign scan from another body, then the attack will likely be discovered right away. This is because (1) the parameters of the scan (location, number of slices, radiation exposure,...) will not match the technician's settings which change for every scan, (2) the doctor treating the patient (e.g., oncologist) can tell that the patient's scan does not match due to body figure, BMI, gender, and age (not to mention head scans contain a face), and (3) the attacker will not be able to repeat the attack when an additional/follow-up scan is performed (e.g., when tracking cancer growth). This is because, aside from the different parameters, the body will not be in the same precise location, raising suspicion around the duplicate imagery.
	
	%-------------------------------------------------------------------------------
	\section{Countermeasures}\label{sec:countermeasures}
	%-------------------------------------------------------------------------------
	The tampering of DICOM medical files is a well-known concern. In the section we provide a brief overview of solutions for preventing and detecting this attack.
	
	\subsection{Prevention}
	To mitigate this threat, administrators should secure both the data-in-motion (DiM) and the data-at-rest (DaR). To secure data-in-motion, admins should enable encryption between the hosts in their PACS network using proper SSL certificates. This may seem trivial, but after discovering this flaw in the hospital we pen-tested, we turned to the PACS software provider for comment. The company, with over 2000 installations worldwide, confirmed to us that their hospitals do not enable encryption in their PACS because ``it is not common practice''. We were also told that some of the PACS don't support encryption at all.\footnote{See \cite{zetter:online} for further comments.}
	To secure the DaR, servers and anti-virus software on modality and radiologist workstations should be kept up to date, and admins should also limit the exposure which their PACS server has to the Internet. 
	
	\subsection{Detection}
	
	The best way to detect this attack is to have the scanner sign each scan with a digital signature. The DICOM image file standard already allows users to store signatures within the file's data structure \cite{cao2003medical,dicomsecStandard}. However, although some PACS software providers offer this feature, we have not seen it in use within a PACS. If enabled, admins should check that valid certificates are being used and that the radiologists' viewing applications are indeed verifying the signatures. 
	
	%
	
	%To secure data-in-motion, admins should enable encryption between the hosts in their PACS network. This may seem trivial, but aside from the hospital we surveyed, a large PACS software provider with over 2000 installations worldwide, confirmed to us that their hospitals do not enable encryption.
	
	Another method for detecting this attack is digital watermarking (DW). A DW is a hidden signal embedded into an image such that tampering corrupts the signal and thus indicates a loss of integrity. For medical images, this subject has been researched in depth \cite{Singh2017} and can provide a means for localizing changes in a tampered image. However, we did not find any medical devices or products which implement DW techniques. This may be due to the fact that they add noise to images which may harm the medical analysis. 
	
	Tampered images can also be detected with machine learning. In the supervised setting (where models are trained on examples of tampered images) the authors in \cite{ghoneim2018medical} propose detection by (1) extracting a scan's noise pattern using a Wiener filter, then (2) applying a multi-resolution regression filter on the noise, and then (3) executing an SVM and ELM together via a Bayesian Sum Rule model. Many domain specific methods exist for detecting images tampered by GANs (e.g., images/videos of faces \cite{rossler2019faceforensics++,matern2019exploiting,tariq2018detecting}). However, the supervised approach in \cite{cozzolino2018forensictransfer} is more suitable for detecting our attack since it is domain generic.
	
	Several approaches have been proposed for unsupervised setting as well. These approaches attempt to detect anomalies (inconsistencies) within the tampered images. To detect these inconsistencies, researchers have considered JPEG blocks, signal processing, and compression/resampling artifacts \cite{zheng2019survey}. For example, in a recent work the authors trained a Siamese network to predict the probability that a pair of patches from two images have the same EXIF metadata (e.g., focal length and shutter speed) \cite{huh2018fighting}. In \cite{huh2018fighting}, the model is trained using a dataset of real images only. In \cite{cozzolino2018noiseprint}, the authors proposed `noiseprint' which uses a Siamese network to extract the camera's unique noise pattern from an image (PRNU) to find inconsistent areas. In their evaluation, the authors show that they can detect GAN-based inpainting. In \cite{Korus2016TIFS}, the authors proposed three strategies for using PRNU-based tampering localization techniques with multi-scale analysis. Using this method, the authors were able to detect forgeries of all shapes and sizes.
	
	While these countermeasures may apply to CT-GAN in some cases, they do admit some caveats; namely, that (1) medical scans are usually not compressed so compression methods are irrelevant, (2) these methods were tested on 2D images and not 3D volumetric imagery, and (3) CT/MR imaging systems produce very different noise patterns than standard cameras. For example, we found that the PRNU method in \cite{Korus2016TIFS} does not work out-of-the-box on our tampered CT scans. This is because the noise patterns of CT images are altered by a radon transform used to construct the image. As future work, we plan to research how these techniques can be applied to detecting attacks such as CT-GAN.

	%medical images do not have compression artifacts
	%need to be validated on 3d volumetric imagery
	%noise different in CT
	%Deepfake detection relaies nmosty on video. Must check if works the same way on sequece so slices.

	%An additional way to verify image integrity is to perform tamper detection. 

	%Another approach to tamper detection is to measuer PNR. [][][] We tested and the results are...  we also test our own method 'using the discermreinator... results are...

	%-------------------------------------------------------------------------------
	\vspace{-.5em}
	\section{Conclusion}\label{sec:conclusion}
	%-------------------------------------------------------------------------------
	
	In this paper we introduced the possibility of an attacker modifying 3D medical imagery using deep learning. We explained the motivations for this attack, discussed the attack vectors (demonstrating one of them), and presented a manipulation framework (CT-GAN) which can be executed by a malware autonomously. As a case study, we demonstrated how an attacker can use this approach to inject or remove lung cancer from full resolution 3D CT scans using free medical imagery from the Internet. We also evaluated the attack and found that CT-GAN can fool both humans and machines: radiologists and state-of-the-art AI. This paper also demonstrates how we should be wary of closed world assumptions: both human experts and advanced AI can be fooled if they fully trust their observations. 
	
	%-------------------------------------------------------------------------------
	%\section*{Acknowledgments}
	%-------------------------------------------------------------------------------
	
	%CT scanner in Fig X licensed an created by Vectorpouch - Freepik.com
	%The authors would like to thank Roman Tsirkin for his invaluable help in the pen-testing.
	
	%-------------------------------------------------------------------------------
	%\section*{Availability}
	%-------------------------------------------------------------------------------
	%A video of the pen-test and example injections/removals can be found at \url{https://tinyurl.com/CTGANanom}.
	%The source code and models used in this paper are available online at [redacted for final print].% \url{https://github.com/ymirsky/CT-GAN}.
	%We have also uploaded the datasets (manipulated CT scans) evaluated by the radiologists and AI.
	%-------------------------------------------------------------------------------

	%\begin{spacing}{0.9}
	
	\bibliographystyle{unsrt}
	\bibliography{paper}
	%\end{spacing}

	%\clearpage
	%\section*{Appendix}
	\appendix
	\large
	\noindent \textbf{Appendix}

	%\begin{figure}
	%	\centering
	%	\includegraphics[width=\columnwidth]{equalize_fig.pdf}
	%	\caption{The affect which histogram equalization has on emphasizing the features in a CT scan.}
	%	\label{fig:equalization}
	%\end{figure}
	\begin{figure}[h]
		\centering
		\includegraphics[width=.49\columnwidth]{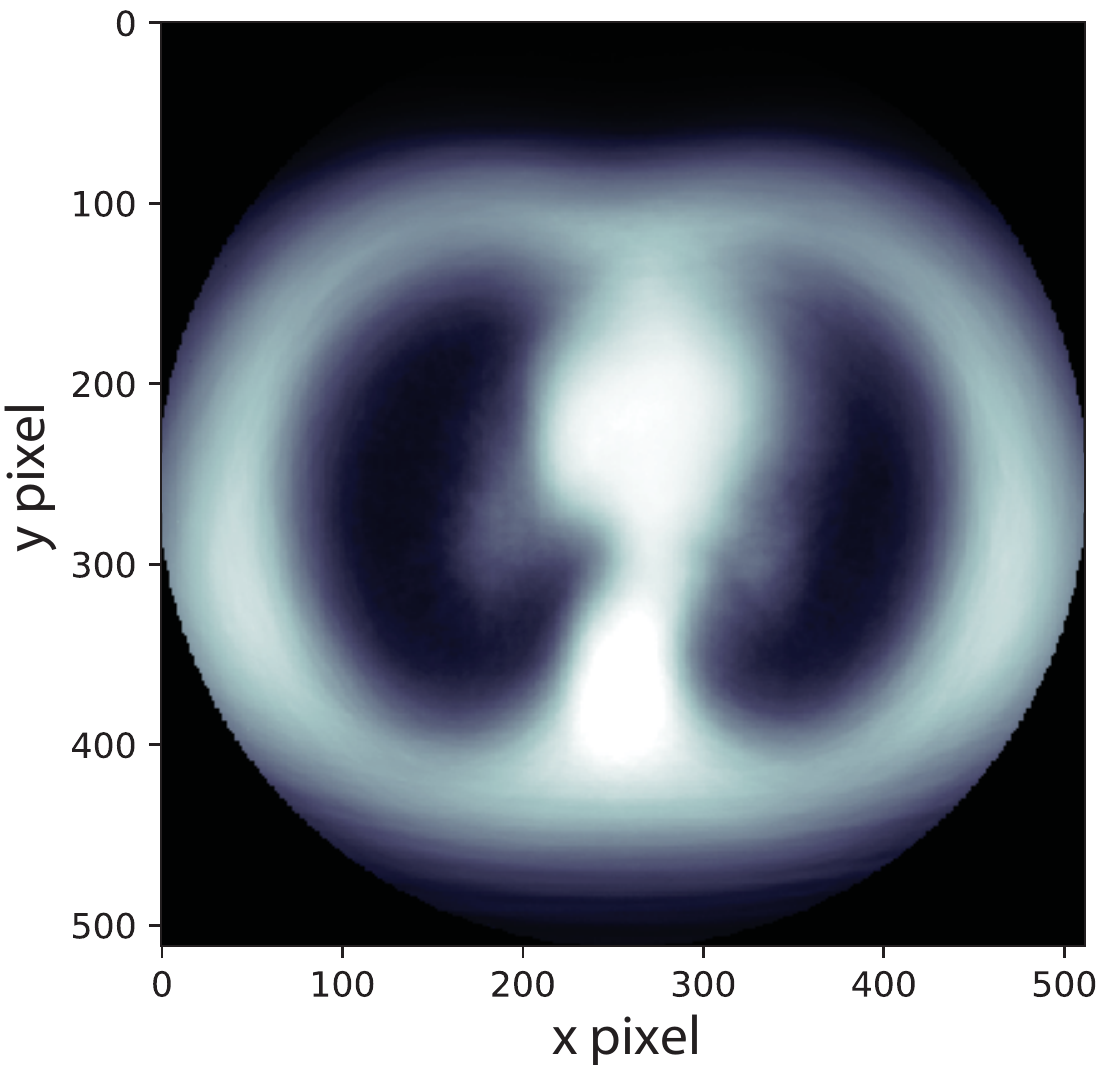}
		\includegraphics[width=.49\columnwidth]{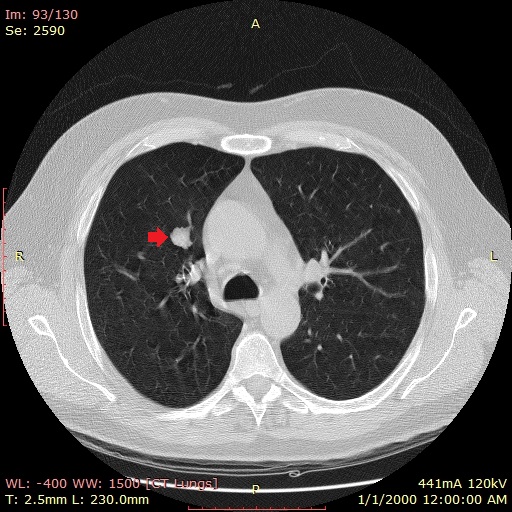}
		\caption{Left: the average of 888 CT scans' middle slices \textit{before} scaling to 1:1:1 ratio (black ares are candidate injection locations). Right: a full slice with an injected nodule.}
		\label{fig:average}
		%\end{figure}
		%\begin{figure}
		\vspace{1em}
		\centering
		\includegraphics[width=\columnwidth]{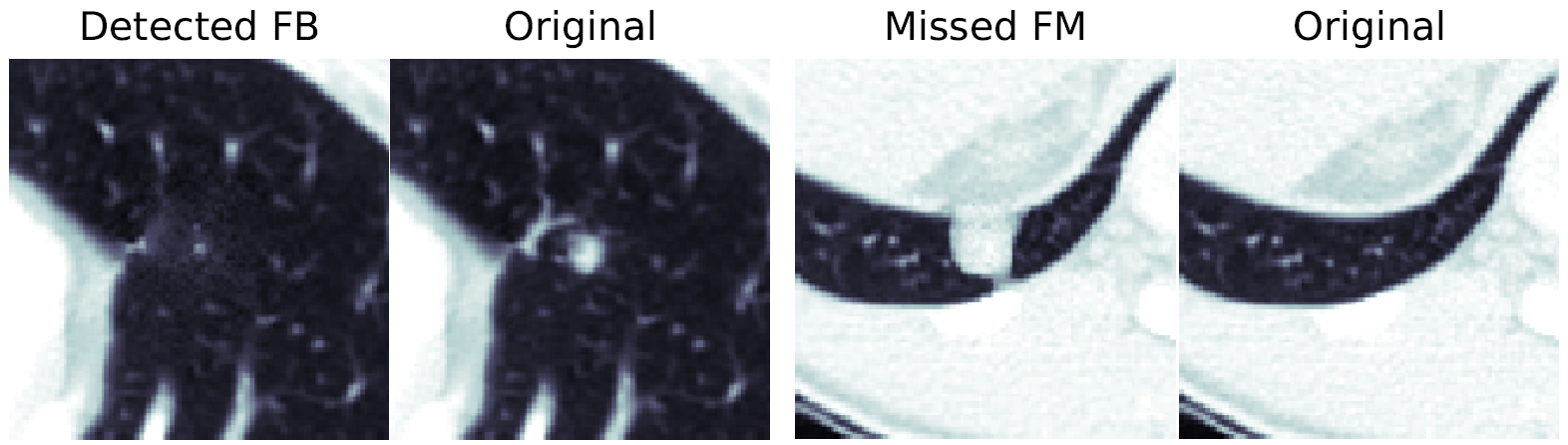}
		\caption{Examples of where the attack failed in the blind trial. Left: a removal (FB) detected as `ground-glass' cancer due to too much additive noise. Right: an injection missed due to human error.}
		\label{fig:mistakes_exp1}
		%\end{figure}
		%\begin{figure}
		\centering
		\includegraphics[width=\columnwidth]{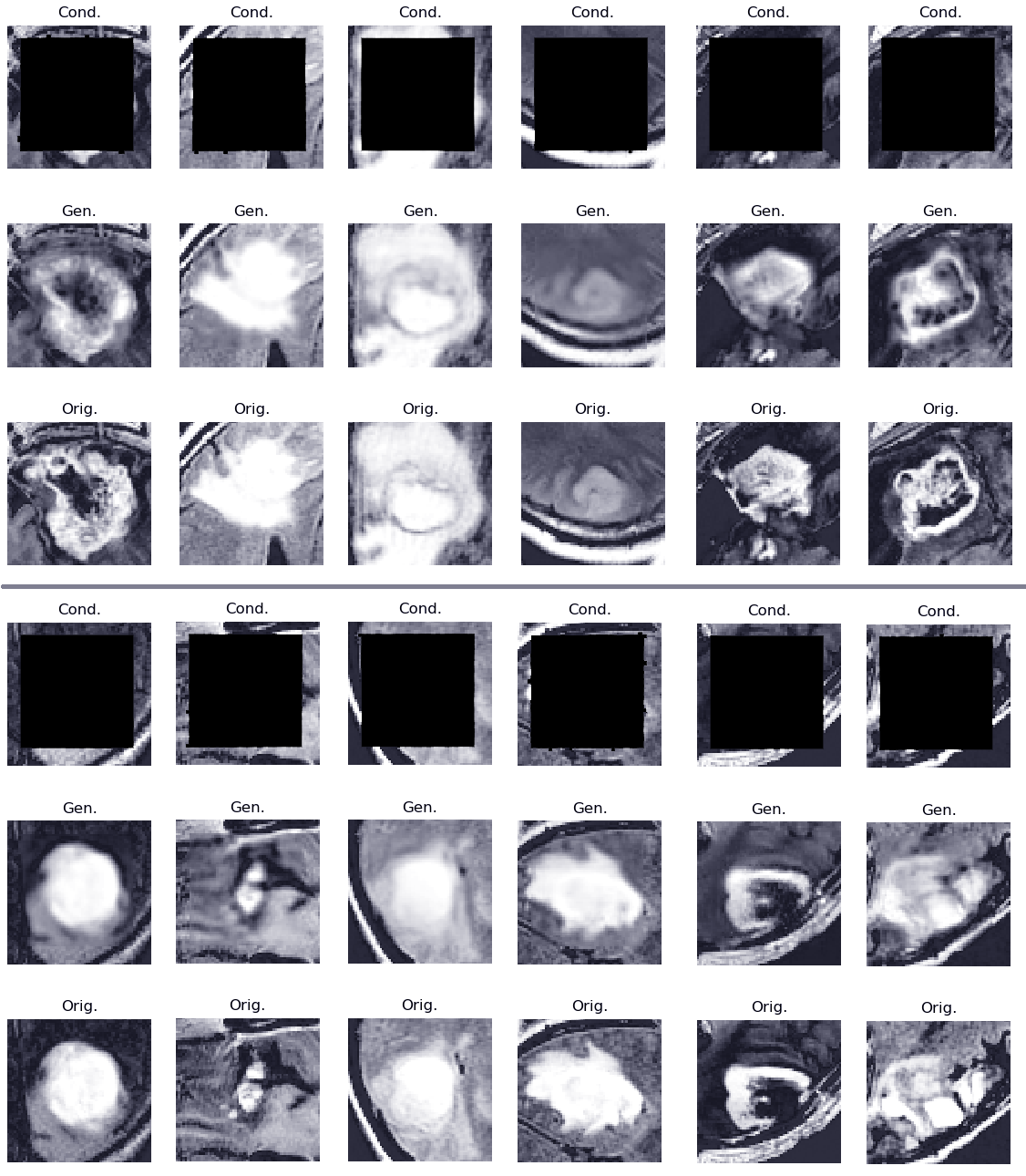}
		\caption{CT-GAN used to inject brain tumors into MRIs of healthy brains. Similar to Fig. \ref{fig:train_prog}, Top: context, middle: in-painted result, and bottom: ground-truth. Showing one slice in a 64x64x16 cuboid.}
		\label{fig:braintumors}
	\end{figure}
	
	\begin{figure*}[t]
		\centering
		\includegraphics[width=1\textwidth]{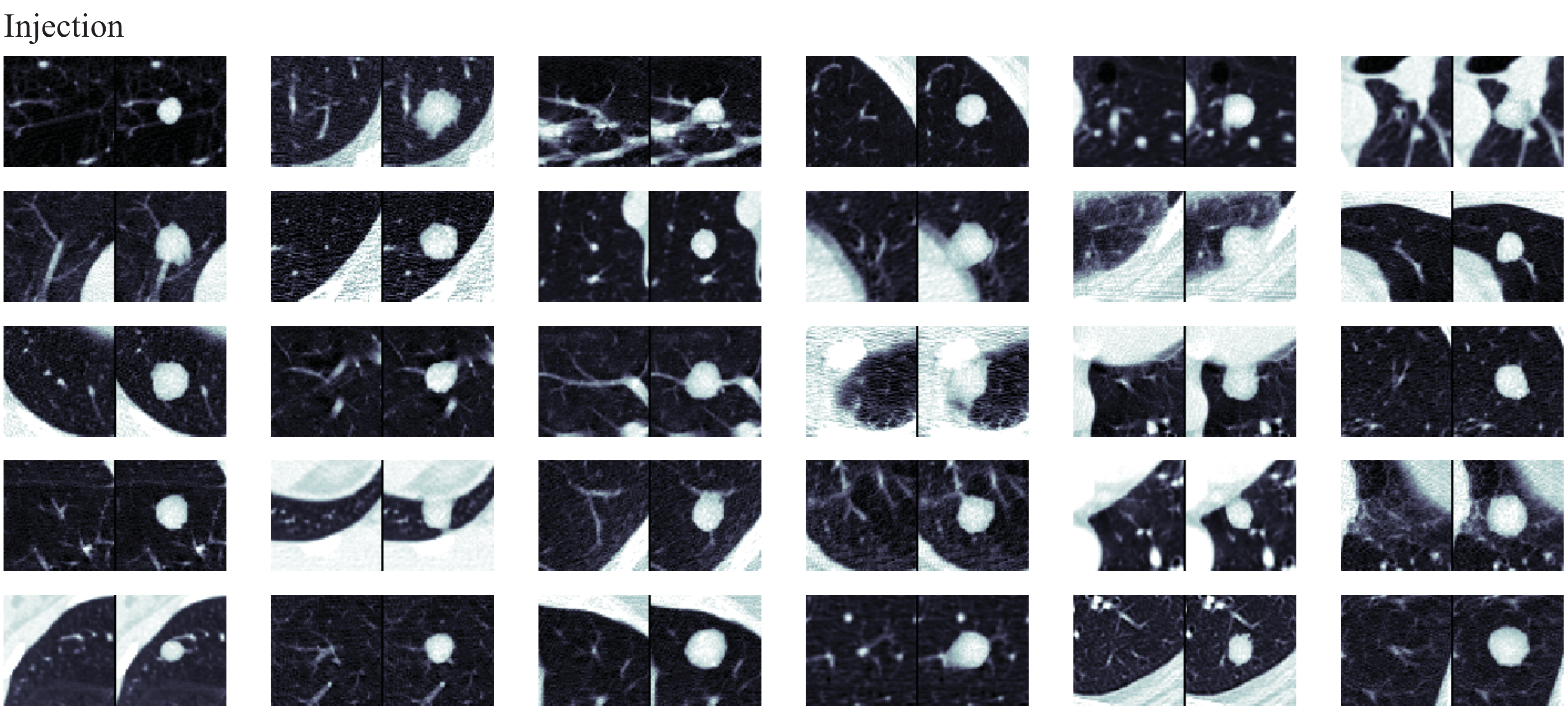}
		\includegraphics[width=1\textwidth]{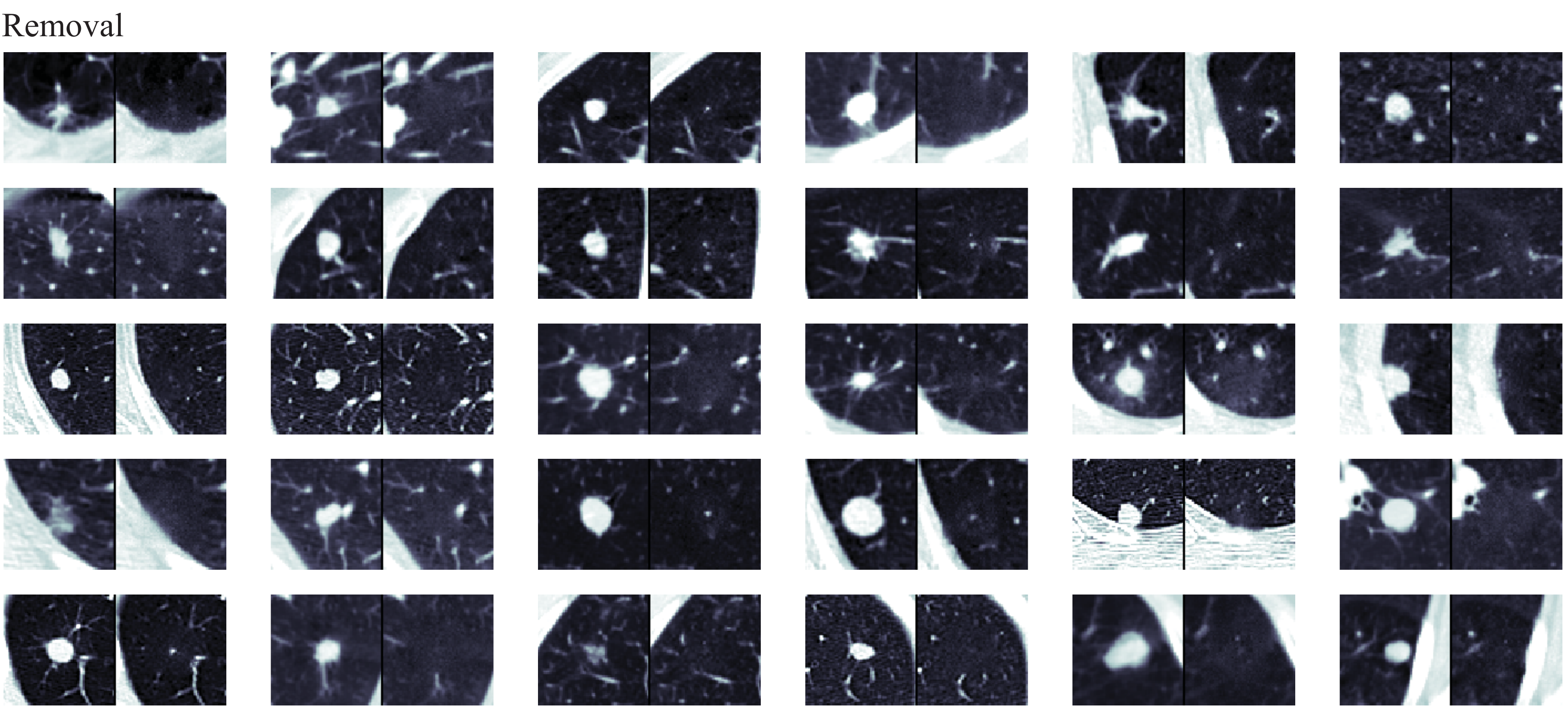}
		\caption{Samples of injected (top) and removed (bottom) pulmonary nodules. For each image, the left side is before tampering and the right side is after. Note, only the middle 2D slice is shown and the images are scaled to different ratios (the source scan).}
		\label{fig:various_examples_extra}
		\vspace{-1em}
	\end{figure*}
	\begin{figure*}[t]
		\centering
		\includegraphics[width=.48\textwidth]{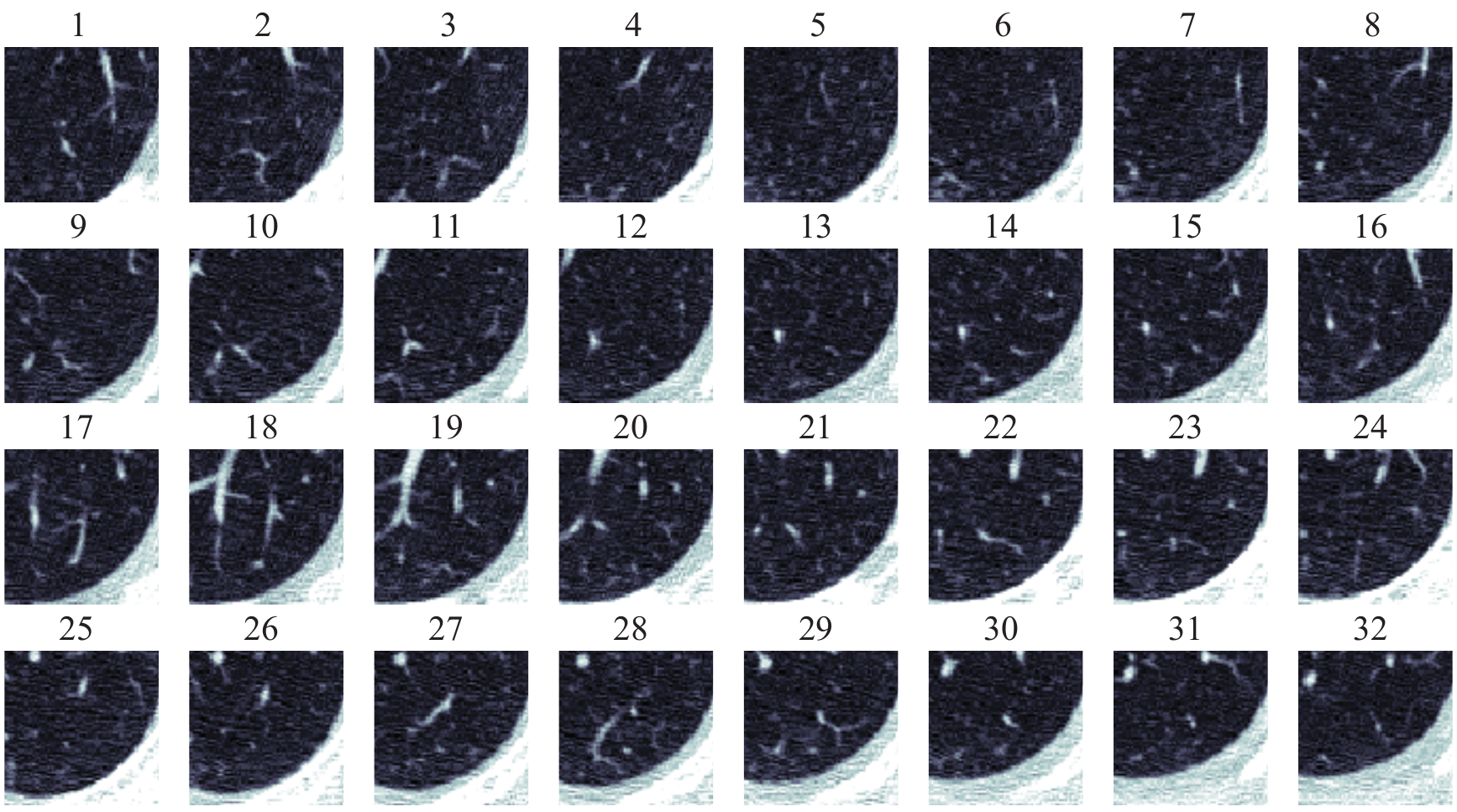}\hspace{1.5em}
		\includegraphics[width=.48\textwidth]{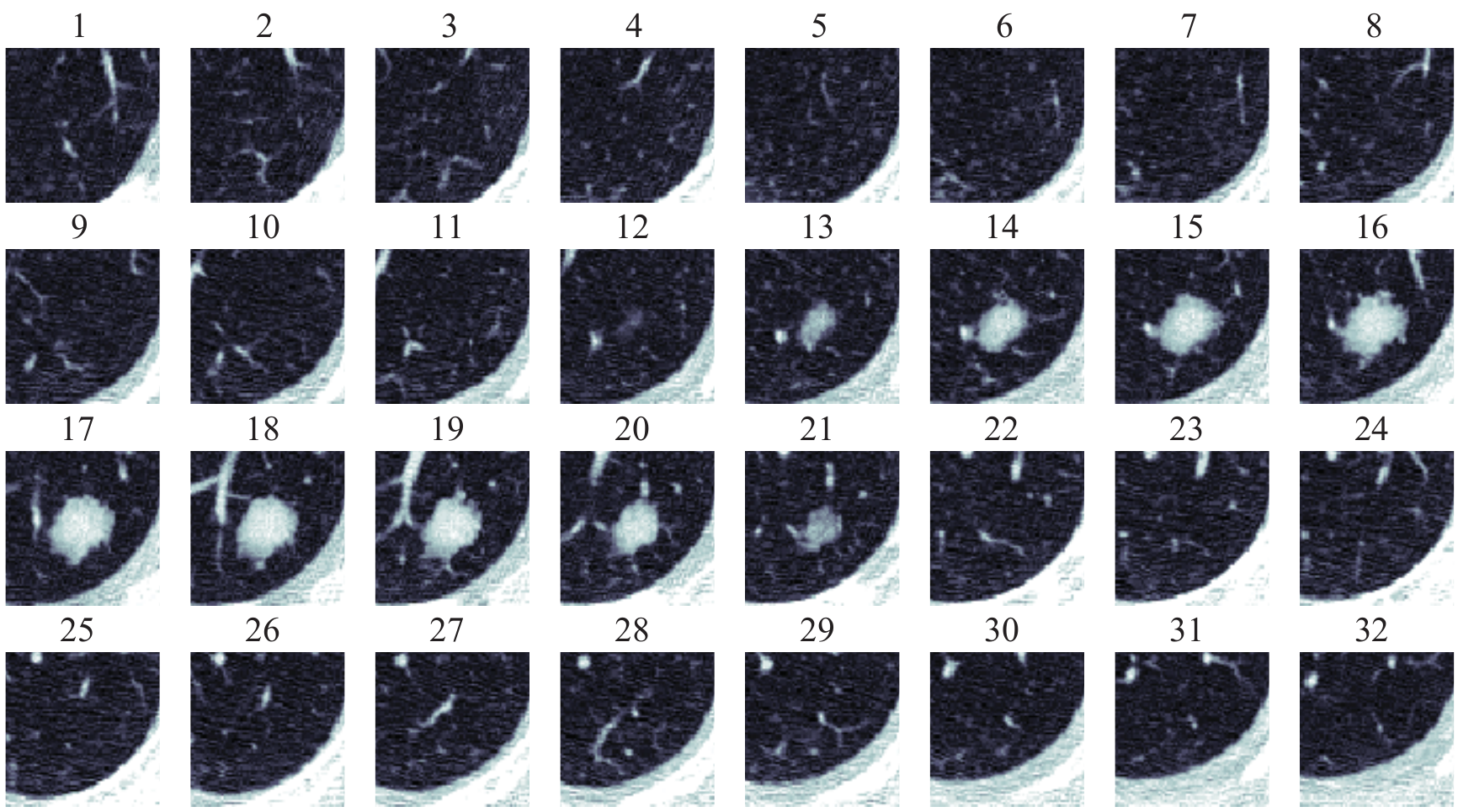}
		
		\caption{All 32 slices from a sample injection before (left) and after (right) tampering with the CT scan.}
		\label{fig:inj_slices_extra}
	\end{figure*}
\end{document}